\documentclass[openacc]{rstransa}

\usepackage[percent]{overpic}

\jname{rsta} 
\Journal{Phil. Trans. R. Soc}

\begin{document}

\title{The 67P/Churyumov-Gerasimenko observation campaign in support of the Rosetta mission}

\author{
C.   Snodgrass$^{1}$
et al. (see below)
}

\address{$^{1}$ School of Physical Sciences, The Open University, Milton Keynes, MK7 6AA, UK\\
}

\subject{Solar System, Observational Astronomy}

\keywords{Comet 67P/Churyumov-Gerasimenko, Rosetta}

\corres{Colin Snodgrass\\
\email{colin.snodgrass@open.ac.uk}}

\begin{abstract}
We present a summary of the campaign of remote observations that supported the European Space Agency's Rosetta mission. Telescopes across the globe (and in space) followed comet 67P/Churyumov-Gerasimenko from before Rosetta's arrival until nearly the end of mission in September 2016. These provided essential data for mission planning, large-scale context information for the coma and tails beyond the spacecraft, and a way to directly compare 67P with other comets. The observations revealed 67P to be a relatively `well behaved' comet, typical of Jupiter family comets and with activity patterns that repeat from orbit-to-orbit. Comparison between this large collection of telescopic observations and the in situ results from Rosetta will allow us to better understand comet coma chemistry and structure. This work is just beginning as the mission ends -- in this paper we present a summary of the ground-based observations and early results, and point to many questions that will be addressed in future studies.
\end{abstract}


\begin{fmtext}
{\bf Co-Authors:} 
M. F.   A'Hearn$^{2}$,
F.   Aceituno$^{3}$,
V.   Afanasiev$^{4}$,
S.   Bagnulo$^{5}$,
J.   Bauer$^{6}$,
G.   Bergond$^{7}$,
S.   Besse$^{8}$,
N.~Biver$^{9}$,
D.   Bodewits$^{2}$,
H.   Boehnhardt$^{10}$,
B. P.   Bonev$^{11}$,
G.~Borisov$^{5,12}$,
B.   Carry$^{13,14}$,
V.   Casanova$^{3}$,
A.   Cochran$^{15}$,
B.~C.~Conn$^{16,17}$,
B.   Davidsson$^{6}$,
J. K.   Davies$^{18}$,
J.~de~Le\'on$^{19,20}$,
E.~de~Mooij$^{21}$,
M.   de~Val-Borro$^{22,23,24}$,
M.   Delacruz$^{25}$,
M. A.   DiSanti$^{23}$,
J. E.   Drew$^{26}$,
R.   Duffard$^{3}$,
N.~J.~T.~Edberg$^{27}$,
S.~Faggi$^{28}$,
L.~Feaga$^{2}$,
A.   Fitzsimmons$^{21}$,
H.   Fujiwara$^{29}$,
E.~L.~Gibb$^{30}$,
M.   Gillon$^{31}$,
S. F.   Green$^{1}$,
A.   Guijarro$^{7}$,
A.   Guilbert-Lepoutre$^{32}$,
P. J.   Guti{\'e}rrez$^{3}$,
E.~Hadamcik$^{33}$,
O.   Hainaut$^{34}$,
S.   Haque$^{35}$,
R.   Hedrosa$^{7}$,
D.~Hines$^{36}$,
U.   Hopp$^{37}$,
F.~Hoyo$^{7}$,
D.   Hutsem{\'e}kers$^{31}$,
M.   Hyland$^{21}$,
O.   Ivanova$^{38}$,
E.   Jehin$^{31}$,
G. H.   Jones$^{39,40}$,
J.~V.~Keane$^{25}$,
M.~S.~P.~Kelley$^{2}$,
N.   Kiselev$^{41}$,
J.   Kleyna$^{25}$,
M.   Kluge$^{37}$,
M.~M.~Knight$^{2}$,
R.~Kokotanekova$^{10,1}$,
D.~Koschny$^{42}$,
E.   Kramer$^{6}$,
J.~J.~L{\'o}pez-Moreno$^{3}$,
P.   Lacerda$^{21}$,
L. M.   Lara$^{3}$,
J.   Lasue$^{43}$,
~~~~~~~~~~~~~~~~~~~~~~~~~
\end{fmtext}


\maketitle

\noindent
H.~J.~Lehto$^{44}$,
A. C.   Levasseur-Regourd$^{33}$,
J.   Licandro$^{19,20}$,
Z. Y.   Lin$^{45}$,
T.   Lister$^{46}$,
S.~C.~Lowry$^{47}$,
A.   Mainzer$^{6}$,
J.   Manfroid$^{31}$,
J.   Marchant$^{48}$,
A. J.   McKay$^{15}$,
A.   McNeill$^{21}$,
K. J.   Meech$^{25}$,
M.~Micheli$^{49}$,
I.   Mohammed$^{50}$,
M.   Mongui\'o$^{26}$,
F.   Moreno$^{3}$,
O.   Mu{\~n}oz$^{3}$,
M. J.   Mumma$^{23}$,
P.~Nikolov$^{12}$,
C.   Opitom$^{31}$,
J. L.    Ortiz$^{3}$,
L.   Paganini$^{23}$,
M.   Pajuelo$^{14,51}$,
F. J.   Pozuelos$^{3}$,
S.~Protopapa$^{2}$,
T.   Pursimo$^{52}$,
B.   Rajkumar$^{35}$,
Y.   Ramanjooloo$^{25}$,
E.   Ramos$^{7}$,
C.   Ries$^{37}$,
A.~Riffeser$^{37}$,
V.   Rosenbush$^{41}$,
P.   Rousselot$^{53}$,
E. L.   Ryan$^{54}$,
P.   Santos-Sanz$^{3}$,
D. G.   Schleicher$^{55}$,
M.   Schmidt$^{37}$,
R.   Schulz$^{56}$,
A. K.   Sen$^{57}$,
A.   Somero$^{44}$,
A.   Sota$^{3}$,
A.   Stinson$^{5}$,
J.   Sunshine$^{2}$,
A.~Thompson$^{21}$,
G. P.   Tozzi$^{28}$,
C.   Tubiana$^{10}$,
G. L.   Villanueva$^{23}$,
X.   Wang$^{58,59}$,
D. H.   Wooden$^{60}$,
M.   Yagi$^{61}$,
B.   Yang$^{62}$,
B.   Zaprudin$^{44}$
and T. J.   Zegmott$^{47}$

{\scriptsize
\noindent
$^{2}$ Department of Astronomy, University of Maryland, College Park, MD 20742-2421, USA\\
$^{3}$ Instituto de Astrof\'{i}sica de Andaluc\'{i}a, CSIC, Glorieta de la Astronom\'{i}a s/n, 18008 Granada, Spain\\
$^{4}$ Special Astrophysical Observatory, Russian Academy of Sciences, Nizhniy Arkhyz, Russia\\
$^{5}$ Armagh Observatory, College Hill, Armagh, BT61 9DG, Northern Ireland, UK\\
$^{6}$ Jet Propulsion Laboratory, M/S 183-301, 4800 Oak Grove Drive, Pasadena CA 91109, USA\\
$^{7}$ Centro Astron{\'o}mico Hispano-Alem{\'a}n, Calar Alto, CSIC-MPG, Sierra de los Filabres-04550 G{\'e}rgal (Almer{\'i}a), Spain\\
$^{8}$ ESA/ESAC, PO Box 78, 28691 Villanueva de la Ca\~nada, Spain\\
$^{9}$ LESIA, Observatoire de Paris, CNRS, UPMC Univ Paris 06, Univ. Paris-Diderot, 5 Place J. Janssen,  92195 Meudon Pricipal Cedex, France\\
$^{10}$ Max-Planck-Institut f\"{u}r Sonnensystemforschung, Justus-von-Liebig-Weg 3, 37077 G\"{o}ttingen\\
$^{11}$ Department of Physics, American University, 4400 Massachusetts Ave NW, Washington, DC 20016, USA\\
$^{12}$ Institute of Astronomy and National Astronomical Observatory, 72, Tsarigradsko Chauss\'{e}e Blvd., BG-1784, Sofia, Bulgaria\\
$^{13}$ Universit{\'e} C{\^o}te d'Azur, OCA, CNRS, Lagrange, France \\
$^{14}$ IMCCE, Observatoire de Paris, PSL Research University, CNRS, Sorbonne Universit\'es, UPMC Univ Paris 06, Univ. Lille, France\\
$^{15}$ University of Texas Austin/McDonald Observatory, 1 University Station, Austin, TX 78712, USA\\
$^{16}$ Research School of Astronomy and Astrophysics, The Australian National University, Canberra ACT, Australia\\
$^{17}$ Gemini Observatory, Recinto AURA, Colina El Pino s/n, Casilla 603, La Serena, Chile\\
$^{18}$ The UK Astronomy Technology Centre, Royal Observatory Edinburgh, Blackford Hill, Edinburgh, EH9 3HJ\\
$^{19}$ Instituto de Astrof\'{\i}sica de Canarias (IAC), C/V\'{\i}a L\'{a}ctea s/n, 38205 La Laguna, Spain\\
$^{20}$ Departamento de Astrof\'{\i}sica, Universidad de La Laguna, 38206 La Laguna, Tenerife, Spain\\
$^{21}$ Astrophysics Research Centre, School of Mathematics and Physics, Queen's University Belfast, BT7 1NN, UK\\
$^{22}$ Department of Astrophysical Sciences, Princeton University, Princeton, NJ 08544, USA\\
$^{23}$ NASA Goddard Space Flight Center, Astrochemistry Laboratory, Code 691.0, Greenbelt, MD 20771, USA\\
$^{24}$ Department of Physics, The Catholic University of America, Washington, DC 20064, USA\\
$^{25}$ Institute for Astronomy, 2680 Woodlawn Drive, Honolulu, HI 96822, USA\\
$^{26}$ School of Physics, Astronomy \& Mathematics, University of Hertfordshire, College Lane, Hatfield, AL10 9AB, UK\\
$^{27}$ Swedish Institute of Space Physics, {\AA}ngstr{\"o}mlaboratoriet, L{\"a}gerhyddsv{\"a}gen 1, 751 21, Uppsala, Sweden\\
$^{28}$ INAF, Osservatorio Astrofisico di Arcetri, Largo E. Fermi 5, I-50 125 Firenze, Italy\\
$^{29}$ Subaru Telescope, National Astronomical Observatory of Japan, 650 North A'ohoku Place, Hilo, HI 96720, USA\\
$^{30}$ Department of Physics \& Astronomy, University of Missouri, St. Louis, USA\\
$^{31}$ Institut d'Astrophysique et de G{\'e}ophysique, Universit{\'e} de Li{\`e}ge, all{\'e}e du 6 Ao{\^u}t 17, B-4000 Li{\`e}ge, Belgium\\
$^{32}$ Institut UTINAM, UMR 6213 CNRS-Universit\'{e} de Franche Comt\'{e}, Besan\c{c}on, France\\
$^{33}$ LATMOS-IPSL; UPMC (Sorbonne Univ.), BC 102, 4 place Jussieu, 75005 Paris, France\\
$^{34}$ European Southern Observatory,  Karl-Schwarzschild-Strasse 2,   D-85748 Garching bei M\"{u}nchen, Germany\\
$^{35}$ Department of Physics, University of the West Indies, St.  Augustine, Trinidad,  West Indies\\
$^{36}$ Space Telescope Science Institute, Baltimore, MD 21218, USA\\
$^{37}$ University Observatory, Ludwig-Maximilian-University Munich, Scheiner Str. 1, 81679 Munich, Germany\\
$^{38}$ Astronomical Institute of the Slovak Academy of Sciences, SK-05960 Tatransk\'a Lomnica, Slovak Republic\\
$^{39}$ Mullard Space Science Laboratory, University College London, Holmbury St. Mary, Dorking, Surrey RH5 6NT, UK\\
$^{40}$ The Centre for Planetary Sciences at UCL/Birkbeck, Gower Street, London WC1E 6BT, UK\\
$^{41}$ Main Astronomical Observatory of National Academy of Sciences, Kyiv, Ukraine\\
$^{42}$ Research and Scientific Support Department, European Space Agency, 2201 Noordwijk, The Netherlands \\
$^{43}$ Universit{\'e} de Toulouse, UPS-OMP, IRAP, Toulouse, France\\
$^{44}$ Tuorla Observatory, Department of Physics and Astronomy, University of Turku, V\"{a}is\"{a}l\"{a}ntie 20, 21500 Piikki\"{o}, Finland \\
$^{45}$ Graduate Institute of Astronomy, National Central University, No.300 Jhongda Rd., Jhongli city, Taoyuan County, 320 Taiwan\\
$^{46}$ Las Cumbres Observatory Global Telescope Network, 6740 Cortona Drive Ste. 102, Goleta, CA 93117, USA\\
$^{47}$ Centre for Astrophysics and Planetary Science, School of Physical Sciences, The University of Kent, Canterbury, CT2 7NH, UK\\
$^{48}$ Astrophysics Research Institute, Liverpool John Moores University, Liverpool, L3 5RF, UK\\
$^{49}$ ESA SSA-NEO Coordination Centre, Frascati (RM), Italy\\
$^{50}$ Caribbean Institute of Astronomy, Trinidad, West Indies\\
$^{51}$ Secci{\'o}n F{\'i}sica, Departamento de Ciencias, Pontificia Universidad Cat{\'o}lica del Per{\'u}, Apartado 1761, Lima, Per{\'u}\\
$^{52}$ Nordic Optical Telescope, Apartado 474,  E-38700 Santa Cruz de La Palma, Santa Cruz de Tenerife, Spain\\
$^{53}$ University of Franche-Comt\'e, Observatoire des Sciences de l'Univers THETA, Institut UTINAM - UMR CNRS 6213, BP 1615, \\
$^{54}$ SETI Institute, 189 Bernardo Ave. Suite 200, Mountain View, Mountain View, CA 94043, USA\\
$^{55}$ Lowell Observatory, 1400 W. Mars Hill Rd, Flagstaff, AZ 86001, USA\\
$^{56}$ Scientific Support Office, European Space Agency, 2201 AZ Noordwijk, The Netherlands \\
$^{57}$ Department of Physics, Assam University, Silchar 788011, India\\
$^{58}$ Yunnan Observatories, CAS, China, P.O. Box 110, Kunming 650011, Yunnan Province, China\\
$^{59}$ Key laboratory for the structure and evolution of celestial objects, CAS, Kunming 650011, China\\
$^{60}$ NASA Ames Research Center, MS 245-3, Moffett Field, CA 94035-1000, USA\\
$^{61}$ National Astronomical Observatory of Japan, 2-21-1, Osawa, Mitaka, Tokyo, 181-8588, Japan\\
$^{62}$ European Southern Observatory, Alonso de Cordova 3107, Vitacura, Santiago, Chile

}

\section{Introduction}

Comets are mostly studied via telescopes, and while the European Space Agency (ESA)'s Rosetta mission has told us much more about comet 67P/Churyumov-Gerasimenko (hereafter 67P) than remote observation alone could ever achieve, remote observation of the comet is necessary for a number of reasons. Firstly, observations were used to characterise the comet ahead of the spacecraft's arrival, to plan the mission. Secondly, the comet's coma and tails stretch thousands to millions of kilometres, far beyond Rosetta's orbit, so a wider view is necessary to understand the total activity and the large scale context that complements the in situ view from the spacecraft. Finally, telescopic observations allow a comparison between 67P and other comets, the vast majority of which will only ever be astronomical objects and not visited directly. Parallel observations allow Rosetta measurements to provide `ground truth' to compare with the interpretation of observations, allowing various techniques to be tested, and for the lessons from Rosetta to be applied to the wider comet population.

The world-wide campaign of observations of 67P includes most major observatories, and deploys all possible techniques across a wide range of wavelengths, from ultraviolet to radio. Unlike previous comet mission support campaigns (for example those supporting the NASA Deep Impact and EPOXI missions\cite{Meech05,Meech11}), the Rosetta mission and campaign are unique in their long duration -- there is not a single fly-by or impact to observe, but rather the long-term evolution of the comet as it approached and then retreated from the Sun. The campaign has been coordinated via a website\footnote{\url{http://www.rosetta-campaign.net}}, mailing lists, and regular meetings. The coordination largely began with a meeting in London in 2012, sponsored by the European Union FP7 research infrastructure `Europlanet' under its networking activity fund\footnote{\url{http://europlanet-scinet.fi/}}. Further meetings were hosted by the European Southern Observatory (ESO) and ESA (usually as parallel sessions to Rosetta Science Working Team meetings). Europlanet again sponsored a workshop in June 2016, at Schloss Seggau near Graz in Austria, towards the end of the parallel Rosetta observations, where results could be exchanged and further analyses of data planned. In addition to the wide range of observations from professional observatory facilities, a large number of amateur astronomers have collected a significant and useful data set. The amateur campaign was coordinated with support from the NASA Rosetta project office, in parallel with and as part of the main campaign, and is described in detail elsewhere\cite{Padma-paper}.

This paper presents an overview of the observations of 67P, together with a
review of some key results from the observing campaign. These include description of the large-scale morphology of
the comet (section~\ref{sec:morphology}), results from spectroscopy (section~\ref{sec:spec}) and polarimetry (section~\ref{sec:pol}), and estimates of total activity levels (section~\ref{sec:activity}). 
Further detailed studies are ongoing, but some other preliminary results, and discussion on their implications, are
included in section~\ref{sec:discussion}.


\section{Observations}

Prior to its selection as the Rosetta mission target in 2003, comet 67P was not particularly well studied. It was discovered in 1969 and observed at its 1983, 1995 and 2002 perihelion passages as part of  narrowband photometry surveys of comets\cite{Schleicher2006}, and was targeted at larger heliocentric distance (for nucleus observations) in between these\cite{Mueller92,Lowry03}. While the original target of the Rosetta mission, comet 46P/Wirtanen, was studied in detail as the mission was developed\cite{Schulz1999}, the delay in the launch of the mission (due to concerns about the launch vehicle) meant that 67P was selected only a year before Rosetta launched towards it, and the relatively unknown comet suddenly became the target of many observations. 67P was just past its perihelion at the time, and the first observations from ESO constrained gas activity levels via spectroscopy\cite{Schulz2004}, while Hubble Space Telescope (HST) imaging was used to estimate nucleus size, shape and rotation rate information using coma subtraction techniques\cite{Lamy06,Lamy-SSR}. Imaging and polarimetric observations at this and the next perihelion passage (in 2009) were used to constrain the dust activity levels and morphology of the coma\cite{Lara05,Hadamcik10,Lara11,Tozzi11}, including large scale structures\cite{Vincent2013}, to monitor changes in dust properties and produce models of the dust size distribution\cite{Hadamcik10,Agarwal10,Fulle10}. Around the aphelion passage between these, a series of observations were used to pin down nucleus properties\cite{Tubiana08,Tubiana11,Lowry12,Kelley06,Kelley09}.

These observations around a full orbit following selection as the Rosetta target meant that, by 2010, 67P was one of the best characterised Jupiter family comets that had not yet been visited by a spacecraft. An analysis of images taken from archives over all previously observed orbits allowed predictions on total activity to be made\cite{Snodgrass2013}, which were largely confirmed during the 2014-2016 Rosetta mission parallel observations. A summary of all observations obtained through the 2009 perihelion passage (up until the end of 2010) is given in table 2 of ref.~\cite{Snodgrass2013}. Additional archival images from the 1 m Jacobus Kapteyn Telescope (JKT) and 2.5 m Isaac Newton Telescope (INT) on La Palma in April 2003 and February 2004, and the 4 m SOAR in August and September 2007, have subsequently been identified. In addition, there were regular observations during the 2011-2012 aphelion passage from ESO telescopes, despite the comet being both faint and located in the direction of the crowded star fields towards the galactic centre. 

\begin{figure}
\begin{center}
   \includegraphics[width=0.7\columnwidth]{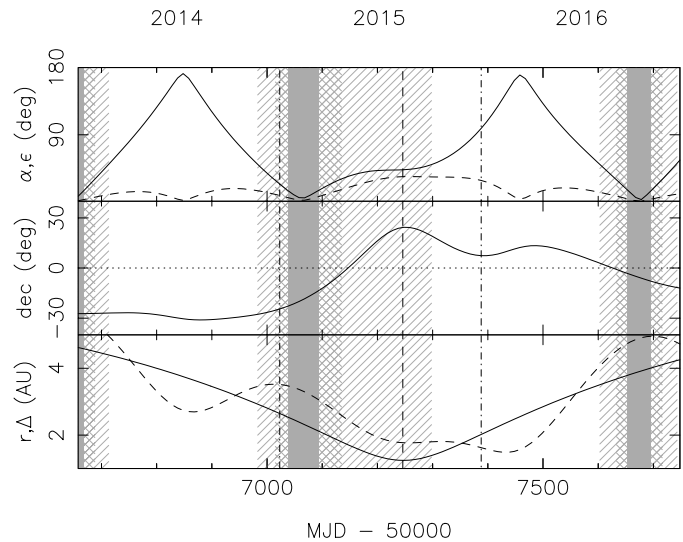} 
\caption{Observability of the comet, as seen from Earth, during the Rosetta mission. The observability of the comet from Earth is shown by hatched, cross-hatched and solid grey areas marking when the solar elongation is less than 50$^\circ$, 30$^\circ$ and 15$^\circ$, respectively. Perihelion (in August 2015) is marked by a vertical dashed line. At that time the comet was 43$^\circ$ from the Sun. Dash-dot vertical lines show the boundaries between the years 2014-2016. Upper panel: Solar elongation $\epsilon$ (solid line) and phase angle $\alpha$ (dashed line); Middle panel: Declination; Lower panel: Heliocentric $r$ (solid line) and geocentric $\Delta$ (dashed line) distances.}
\label{fig:visibility}
\end{center}
\end{figure}

\begin{table}
\caption{Summary table of observations. }
\begin{center}
\tiny
\begin{tabular}{llllll}
\hline
Telescope/Instrument & Technique & Wavelength range & Dates (YY/MM/DD) & ToT & PI \\
\hline
VLT/FORS & IMG & R & 13/04/17 -- 16/06/20 & 97.3 & C Snodgrass \\
VLT/FORS & IMG & R & 13/05/11 -- 13/06/18 & 0.9 & KJ Meech \\
NOT/ALFOSC & IMG & V,R & 13/05/13 -- 16/05/27 & 103.6 & H Lehto \\
NOT/StanCam & IMG & V,R & 14/04/05 -- 16/05/22 & 11.4 & H Lehto \\
VLT/FORS & SPEC & 330-1100 nm & 14/05/07 -- 13/06/18 & 56.1 & C Snodgrass \\
UH88-TEK & IMG & R & 14/06/26 -- 14/06/26 & 2.7 & KJ Meech \\
LOT (1-m) & IMG & BVRI, NB OSIRIS set & 14/06/30 -- 15/11/30 & 50.8 & ZY Lin \\
HST/ACS/WFC & POL & F606W & 14/08/18 -- 16/03/07 & 59.2 & D Hines \\
Gemini S/Flamingos-2 & IMG & J,H,K & 14/09/19 -- 15/06/30 & 5.5 & MM Knight \\
Gemini S/GMOS & IMG & g,r,i,z & 14/09/20 -- 14/11/19 & 4.5 & MM Knight \\
OGS/SDC & IMG & visible & 14/09/21 -- 16/07/04 & 4.6 & D Koschny \\
CFHT/MegaCam & IMG & g,r & 14/10/24 -- 16/05/10 & 0.3 & KJ Meech \\
VLT/XSHOOTER & SPEC & 0.3-2.5 $\mu$m & 14/11/09 -- 14/11/16 & 10.1 & C Snodgrass \\
TRAPPIST 0.6m & IMG & B,V,R,I, CN,C2,BC,GC,RC & 15/04/18 -- 16/06/07 & 22.2 & E Jehin \\
SATU/St Augustine - Tuorla CCD & IMG & R & 15/04/23 -- 15/06/17 & 4.0 & H Lehto \\
ALMA & SPEC & 293-307, 343-355 GHz & 15/05/17 -- 15/09/27 & 5.8 & N Biver \\
VLT/UVES & SPEC & 304-1040 nm & 15/06/24 -- 16/02/10 & 10.0 & E Jehin \\
WHT/ACAM & IMG/SPEC & R,I, g,r,i / 350-940 nm & 15/07/07 -- 16/06/28 & 4.0 & A Fitzsimmons / C Snodgrass \\
TNG/NICS & IMG & J,H,K & 15/07/11 -- 15/12/13 & 5.6 & GP Tozzi / C Snodgrass \\
STELLA/WIFSIP1 & IMG & g,r,i,z & 15/07/18 -- 16/06/08 & 39.7 & C Snodgrass \\
LT/IO:O & IMG & g,r,i,z & 15/07/19 -- 16/06/11 & 22.4 & C Snodgrass \\
IRTF/CSHELL & SPEC & 1-5 $\mu$m & 15/07/26 -- 15/07/31 & 3.9 & L Paganini \\
Gemini N/NIRI & IMG & J,H,K & 15/08/04 -- 16/05/23 & 15.3 & MM Knight \\
LCOGT 2.0m/CCD & IMG & g,r,i,z & 15/08/08 -- 15/09/22 & 1.5 & T Lister \\
2m BNAO-Rozhen/FoReRo2 & IMG & R, NB 387,443,614,642,684 nm & 15/08/11 -- 16/04/28 & 8.1 & G Borisov / P Nikolov \\
CA 2.2m/CAFOS & IMG & R & 15/08/14 -- 16/06/05 & 49.8 & F Moreno \\
CA 3.5m/MOSCA & IMG & R & 15/08/18 -- 15/08/25 & 0.4 & F Moreno \\
TNG/DOLORES & IMG/SPEC & B,V,R / 300-843 nm & 15/08/18 -- 16/06/06 & 16.6 & GP Tozzi / C Snodgrass \\
Lowell 0.8m/NASAcam & IMG & R,CN & 15/08/18 -- 15/12/01 & 15.0 & MM Knight \\
WHT/ISIS & SPEC & 300-1020 nm & 15/08/20 -- 16/04/27 & 7.5 & A Fitzsimmons / C Snodgrass \\
Wendelstein/2m & IMG & g,r,i & 15/08/21 -- 16/05/09 & 94.9 & H Boehnhardt \\
Wendelstein/0.4m & IMG & r,i & 15/08/21 -- 15/11/11 & 29.8 & H Boehnhardt\\
LT/SPRAT & SPEC & 400-800 nm &	15/09/04 -- 16/01/12	& 1.85 & C Snodgrass\\
LT/LOTUS & SPEC & 320-630 nm & 15/09/05 -- 16/01/12 & 2.7 & C Snodgrass \\
Lowell 1.1m/Kron photometer & PHOT & OH,NH,CN,C3,C2,UVC,BC,GC & 15/09/12 -- 15/10/15 & 2.0 & DG Schleicher \\
IRAM-30m/EMIR & SPEC & 3.4-0.97 mm & 15/09/18 -- 15/09/22 & 8.0 & N Biver \\
OSN 1.52m/CCD & IMG & R & 15/09/22 -- 15/11/28 & 13.0 & F Moreno \\
DCT/LMI & IMG & R,r,CN,OH,BC,RC & 15/09/23 -- 16/05/26 & 4.9 & MM Knight / MSP Kelley / D Bodewits \\
GTC/OSIRIS & IMG & r,NB 514,530,704,738,923 nm & 15/09/29 -- 16/02/10 & 5.9 & C Snodgrass \\
INT/IDS & SPEC & 300-610 nm & 15/10/07 -- 15/10/07 & 0.7 & C Snodgrass \\
Gemini N/GNIRS & SPEC & 1-2.5 $\mu$m & 15/10/14 -- 16/01/04 & 2.6 & MM Knight \\
INT/WFC & IMG & B,r,i,z & 15/10/14 -- 16/06/21 & 56.1 & C Snodgrass / A Fitzsimmons / SC Lowry \\
WHT/LIRIS & IMG & J,H,K & 15/10/29 -- 16/01/23 & 3.0 & C Snodgrass \\
6m BTA SAO RAS/SCORPIO-2 & IMG/SPEC/IPOL & g,r / 350-707 nm / R & 15/11/08 -- 16/04/05 & 3.7 & N Kiselev / V Rosenbush \\
Odin sub-mm receivers & SPEC & 0.54 mm & 15/11/09 -- 15/11/12 & 63.8 & N Biver \\
Lijiang (2.4m) & IMG & R, NB OSIRIS set & 15/11/19 -- 16/01/06 & 8.3 & ZY Lin \\
TNG/HARPS-N & SPEC & 383-693 nm & 15/12/09 -- 15/12/09 & 0.3 & C Snodgrass \\
TBL/Narval & SPEC & 370-1000 nm & 15/12/10 -- 15/12/11 & 2.3 & J Lasue \\
OHP 80cm  & IMG & visible & 15/12/11 -- 16/01/08 & 6.0 & E Hadamcik \\
HCT (2m) & IMG & R,I & 15/12/12 -- 15/12/12 & 2.5 & AK Sen \\
WHT/ISIS & IPOL & r & 15/12/18 -- 16/03/11 & 18.0 & C Snodgrass / S Bagnulo \\
NEOWISE & IMG & 3.4,4.6 $\mu$m & 15/12/21 -- 16/05/23 & 0.1 & A Mainzer / J Bauer \\
Keck/HIRES & SPEC & 350-1000 nm & 15/12/26 -- 15/12/27 & 8 & A McKay \\
VLT/FORS & IPOL/PMOS & R, NB 485 nm / 400-950 nm & 16/01/10 -- 16/03/04 & 8.2 & S Bagnulo \\
OSN 0.9m/CCD & IMG & R & 16/01/14 -- 16/01/16 & 3.0 & F Moreno \\
LCOGT 1.0m/CCD & IMG & r & 16/01/30 -- 16/03/06 & 1.3 & T Lister \\
IRTF/SPEX & SPEC & 0.8-2.5 $\mu$m & 16/02/04 -- 16/03/28 & 17.5 & S Protopapa / Y Ramanjooloo \\
Gemini N/GMOS & IMG & g,r,i,z & 16/02/16 -- 16/05/28 & 2.0 & MM Knight \\
VLT/MUSE & SPEC & 465-930 nm & 16/03/03 -- 16/03/07 & 9.0 & A Guilbert-Lepoutre \\
VLT/SINFONI & SPEC & H+K & 16/03/03 -- 16/03/07 & 7.4 & A Guilbert-Lepoutre \\
Subaru/HSC & IMG & HSC-g (480 nm) & 16/03/08 -- 16/03/08 & 1.1 & M Yagi \\
IRTF/MORIS & IMG & r & 16/03/13 -- 16/03/28 & 13.5 & Y Ramanjooloo \\
Spitzer & IMG & 3.6,4.5 $\mu$m & 16/04/08 -- 16/05/08 & 1.3 & MSP Kelley \\
VLT/VIMOS & IMG & R & 16/05/09 -- 16/05/10 & 1.5 & A Fitzsimmons \\
NTT/EFOSC & IMG & r & 16/07/29 -- 16/07/29 & 0.3 & P Lacerda \\
Kepler & IMG & visible & 16/09/08 -- 16/09/20 & 288.0 & C Snodgrass \\
\hline
\end{tabular}
\end{center}
\small
{\it Notes: 
ToT = time on target (hours). Techniques are IMG = imaging, PHOT = photometry, SPEC = spectroscopy, IPOL = imaging polarimetry, PMOS = spectropolarimetry. Filters in letters for standard bands, with lowercase (griz) indicating SDSS type filters and upper case (BVRI) indicating Johnson/Cousins types. NB = narrowband (followed by central wavelengths), some cometary narrowband filters labelled by name (e.g. CN around CN emission band). Wavelength range given for spectroscopy, in typical unit (nm for visible, $\mu$m for near-IR, etc.).
}
\label{tab:obs-summary}
\end{table}%

\begin{figure}
\begin{center}
\includegraphics[width=\columnwidth]{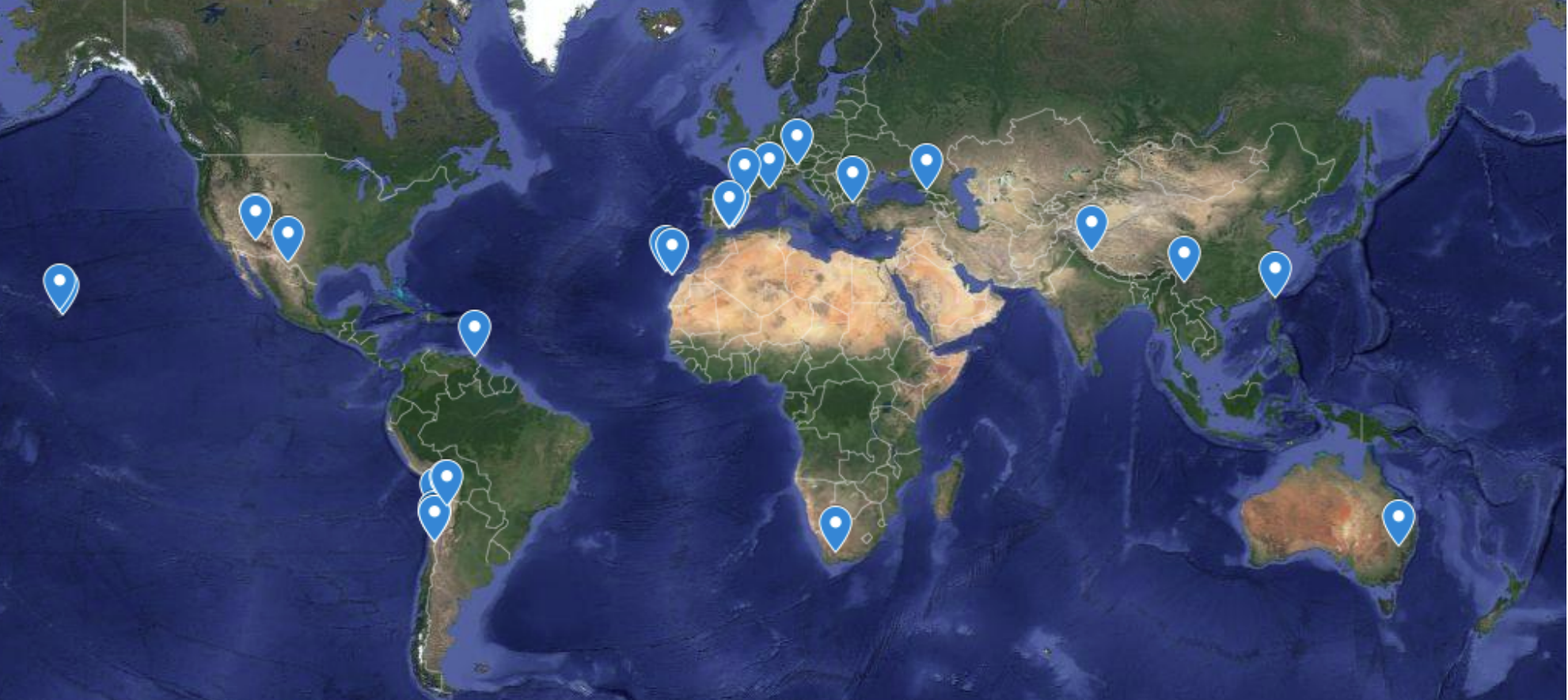}
\caption{Map of locations of contributing observatories.}
\label{fig:map}
\end{center}
\end{figure}

The coordinated campaign of $\sim$parallel observations with Rosetta began in the 2013 observing season, with approximately monthly imaging with the ESO 8~m VLT from April to October, primarily dedicated to astrometric measurements to improve the orbit determination ahead of Rosetta's arrival at the comet in 2014. Further observations with the VLT followed the beginning of detectable activity through 2014, up until the Philae landing in November\cite{Snodgrass2016}, which coincided with the end of the 2014 visibility window from Earth. The comet became brighter as it approached perihelion in August 2015, and was observed by a wide range of facilities through the main visibility window in parallel with Rosetta, which stretched from April 2015 until August 2016. The visibility windows around the Rosetta mission are shown in fig.~\ref{fig:visibility}, and a summary of all observations in the coordinated campaign is given in table~\ref{tab:obs-summary}. More detailed information on the observations can be found on the online log of observations at \url{http://www.rosetta-campaign.net/observations}. The broad geographical spread of participating observatories is illustrated in fig.~\ref{fig:map}. Totalling the (approximate) time on target from each set of observations, we calculate that $\sim$1300 hours of telescope time were dedicated to observing comet 67P during the Rosetta mission.

In 2014 we were mainly limited to larger telescopes, the 8~m VLT and Gemini-S, due to the comet's faintness and Southern declination. There were also observations using the 2.5~m Nordic Optical Telescope (NOT) on the island of La Palma, which has the advantage of being able to point to low elevations, and was therefore one of the only telescopes able to follow the comet over the full observability range from both hemispheres\cite{Zaprudin2015}. In the second quarter of 2015 the comet was briefly visible from Southern sites again, before being a Northern hemisphere target through perihelion, although visibility was limited to a short window before sunrise from any given site. 

Around perihelion robotic telescopes played a large part in the campaign, as these are ideal for obtaining regular short observations\cite{Snodgrass-MNRAS}. One of the key robotic contributors to the campaign was the 0.6~m TRAPPIST telescope at La Silla observatory in Chile\cite{TRAPPIST}, which is dedicated to monitoring comets (and extrasolar planets). A larger robotic facility, the 2~m Liverpool Telescope (LT) on La Palma\cite{Steele2004}, was able to provide spectroscopic monitoring using a new instrument specially designed and commissioned for this observing campaign, LOTUS\cite{lotus}. The LT observations were performed as part of a large International Time Programme across six Canary Island telescopes, which enabled a wide range of observations to be taken with various techniques (broad- and narrow-band imaging, spectroscopy, polarimetry) across the visible and near-IR wavelengths. Near-IR observations were also taken at the NASA IRTF facility on Hawaii and over a long period at the Gemini telescopes, while even longer wavelength observations were possible with the Spitzer and NEOWISE space telescopes in the IR and sub-mm arrays on Earth, including ALMA, near to perihelion. Meanwhile spectroscopic observations continued at the ESO VLT over as wide a time range as possible, despite some very challenging weather in 2016 in Chile. Many other facilities contributed imaging monitoring observations while the comet was relatively bright and well placed in late 2015 and the first half of 2016, with the Wendelstein observatory in Germany\cite{Boehnhardt-MNRAS}, Calar Alto observatory in Spain, ESA's optical ground station on Tenerife, the Lulin observatory in Taiwan and the Lowell Observatory in the USA providing regular and world-wide coverage. 

As Rosetta entered its extended mission in 2016 the comet was increasingly visible all night, although fading as it retreated from the Sun, and was targeted with wide-field imagers, including a serendipitous deep and wide observation with the 8~m Subaru telescope (Hyper Suprime-Cam) on Hawaii, and with integral field unit spectrographs (MUSE and SINFONI at the VLT), to investigate compositional variations across the gas and dust coma. The dust coma was further investigated by measuring its polarisation as the observing geometry changed, using various facilities including the Russian 6~m, the 2~m at Rohzen observatory in Bulgaria, the 4~m William Herschel Telescope on La Palma, the HST, and the VLT. 
Finally, as the Rosetta mission reached its end in September 2016, a last set of remote imaging observations was collected by the NASA Kepler satellite, as the comet happened to be crossing the survey field of this facility in the weeks before the end of mission, after it was no longer visible from Earth.


\section{Large scale morphology}\label{sec:morphology}

   \begin{figure}
   \centering
   \begin{overpic}[width=0.325\columnwidth,trim=111 230 111 230, clip]{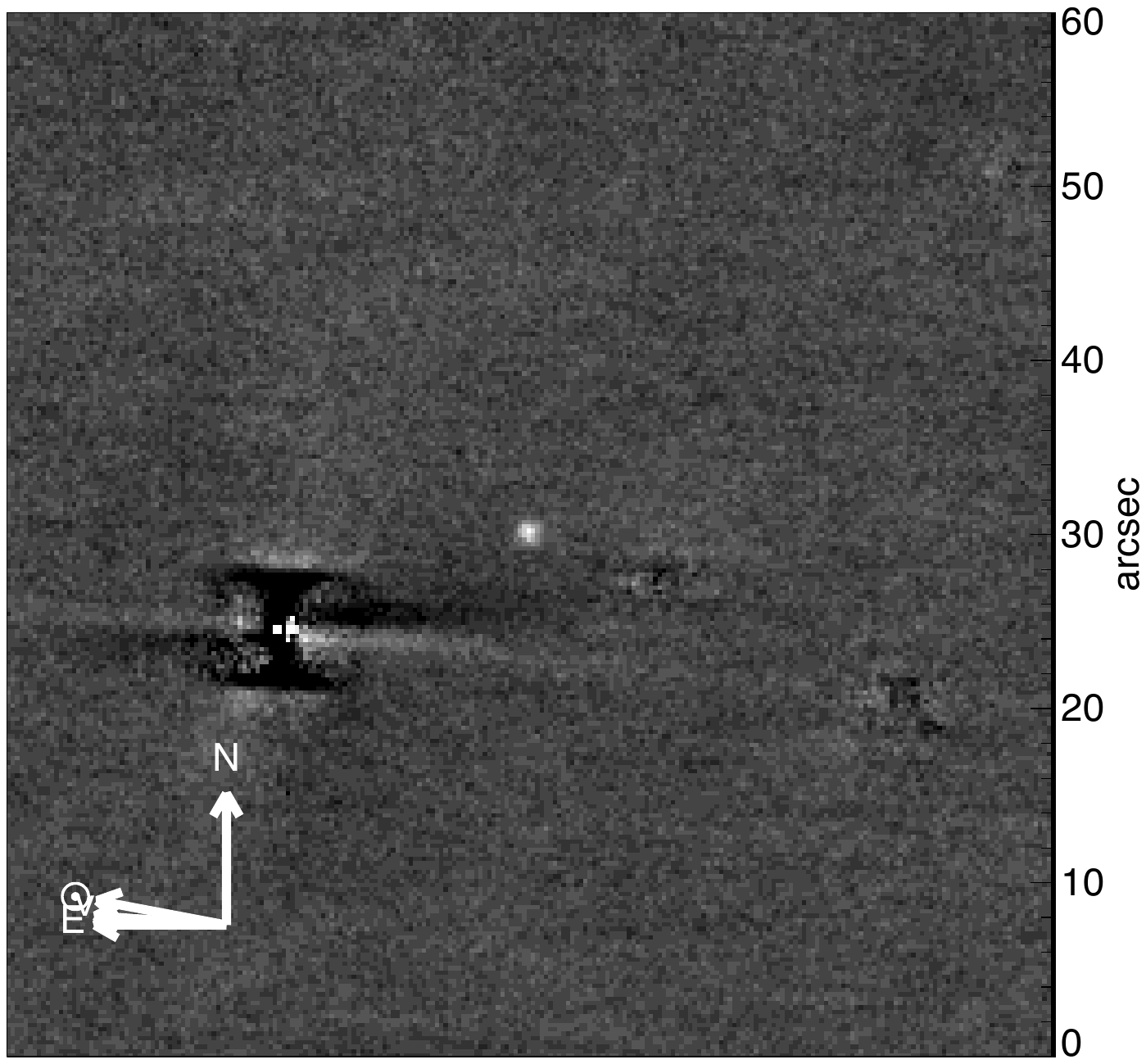}\put(5,85){\textcolor{white}{Feb 2014}}\end{overpic}
   \begin{overpic}[width=0.325\columnwidth,trim=111 230 111 230, clip]{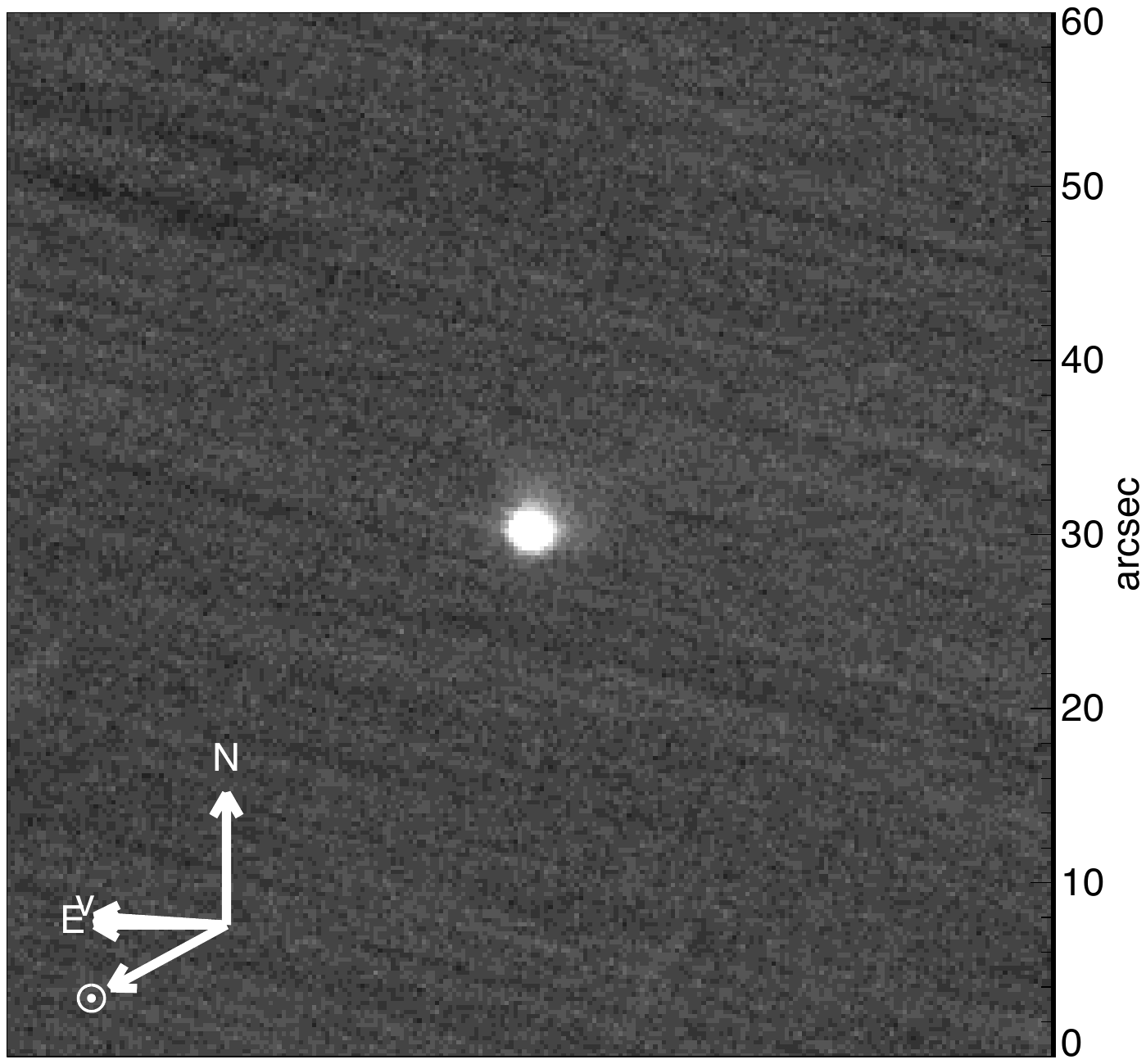}\put(5,85){\textcolor{white}{Jul 2014}}\end{overpic}
   \begin{overpic}[width=0.325\columnwidth,trim=111 230 111 230, clip]{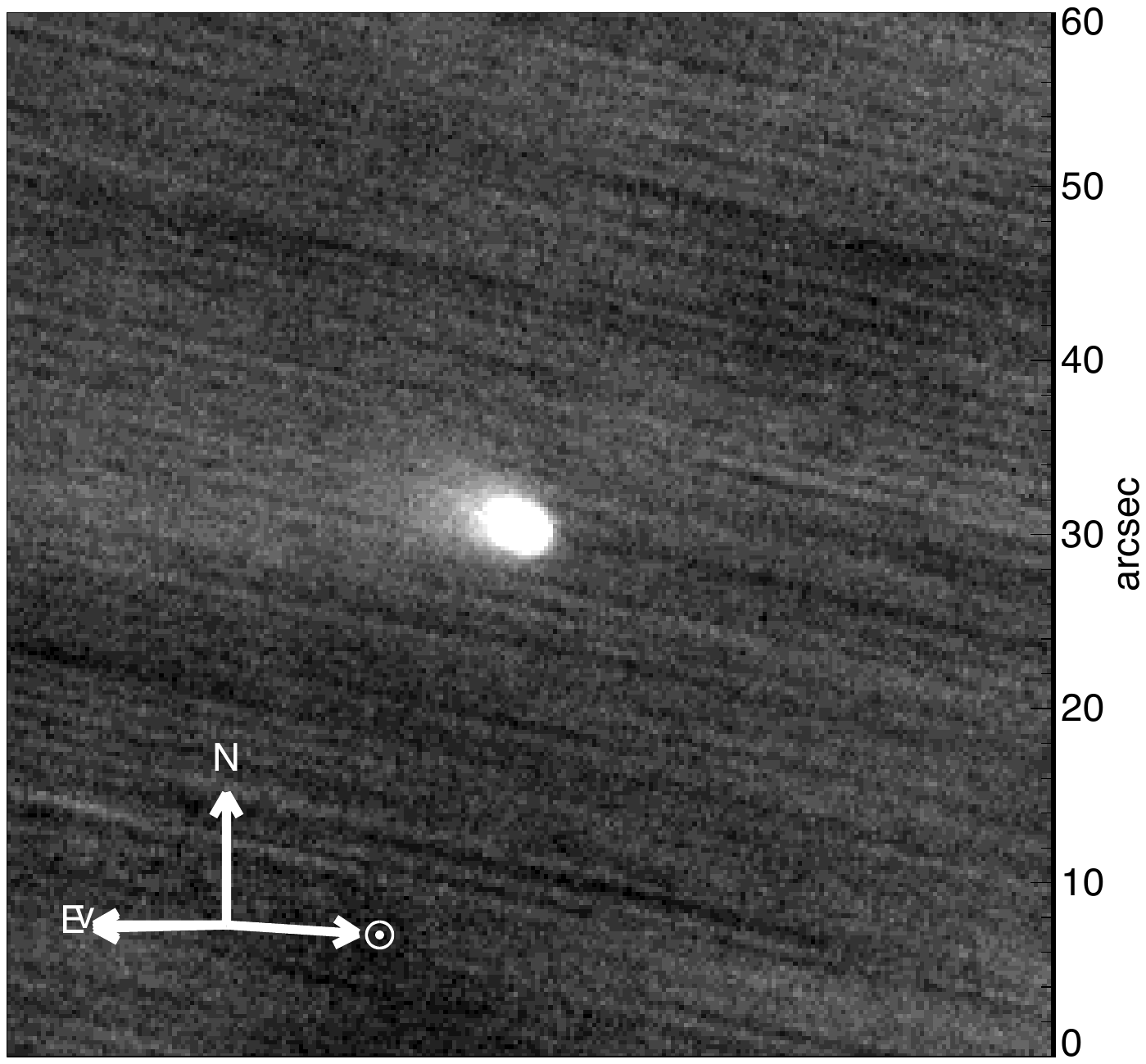}\put(5,85){\textcolor{white}{Oct 2014}}\end{overpic}\\
    \begin{overpic}[width=0.325\columnwidth,trim=111 230 111 230, clip]{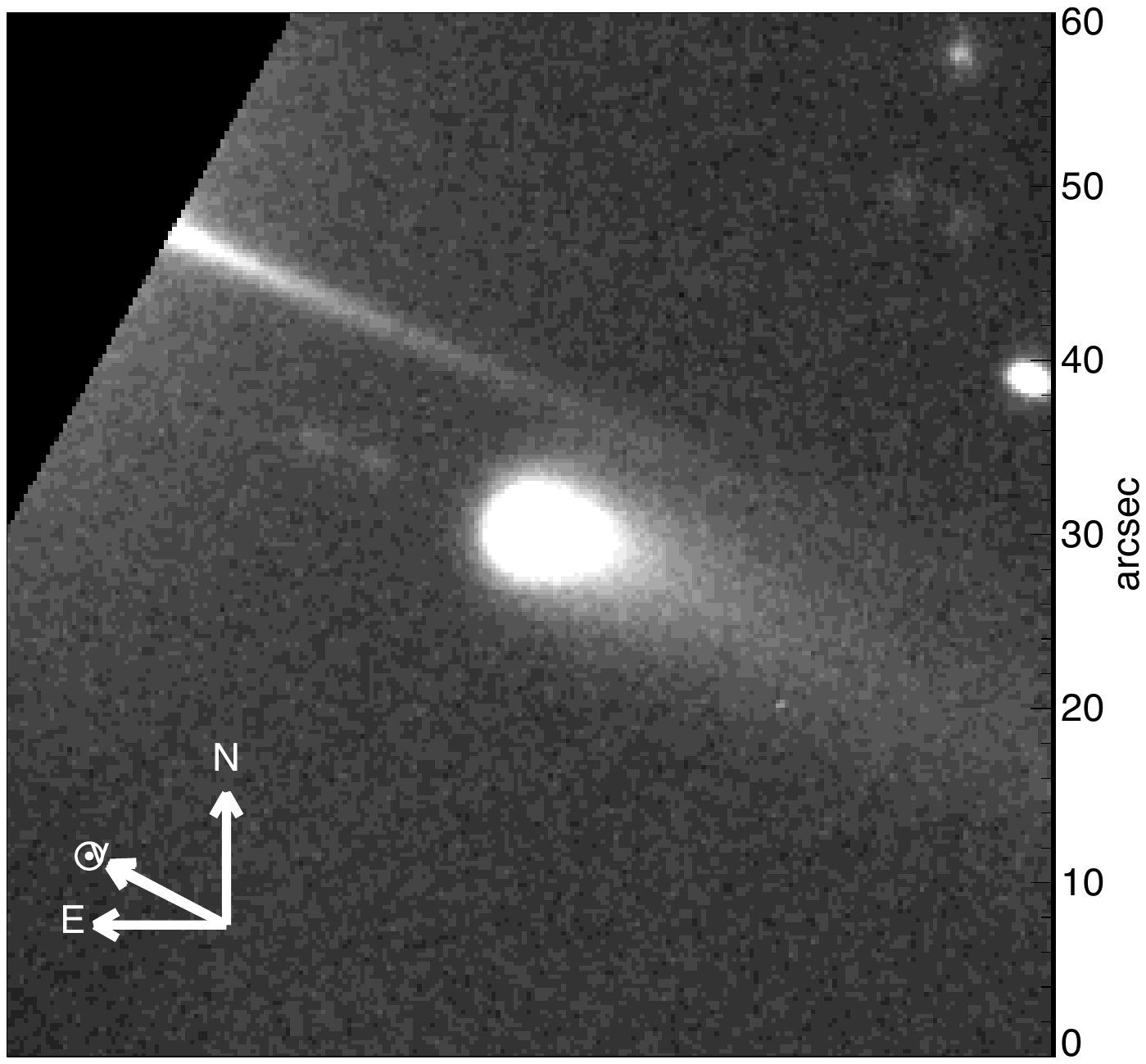}\put(5,85){\textcolor{white}{May 2015}}\end{overpic}
   \begin{overpic}[width=0.325\columnwidth,trim=111 230 111 230, clip]{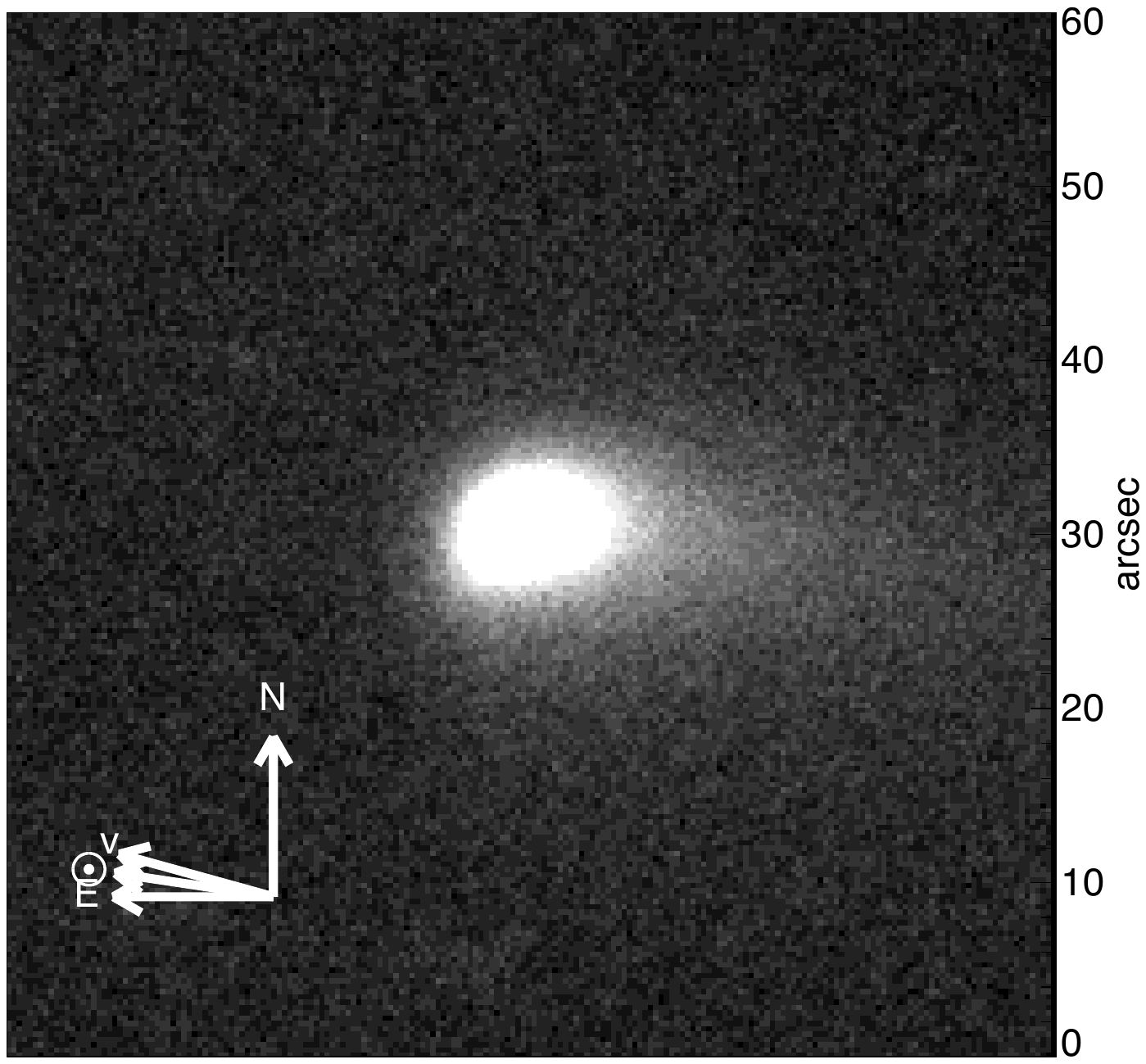}\put(5,85){\textcolor{white}{Jul 2015}}\end{overpic}
   \begin{overpic}[width=0.325\columnwidth,trim=111 230 111 230, clip]{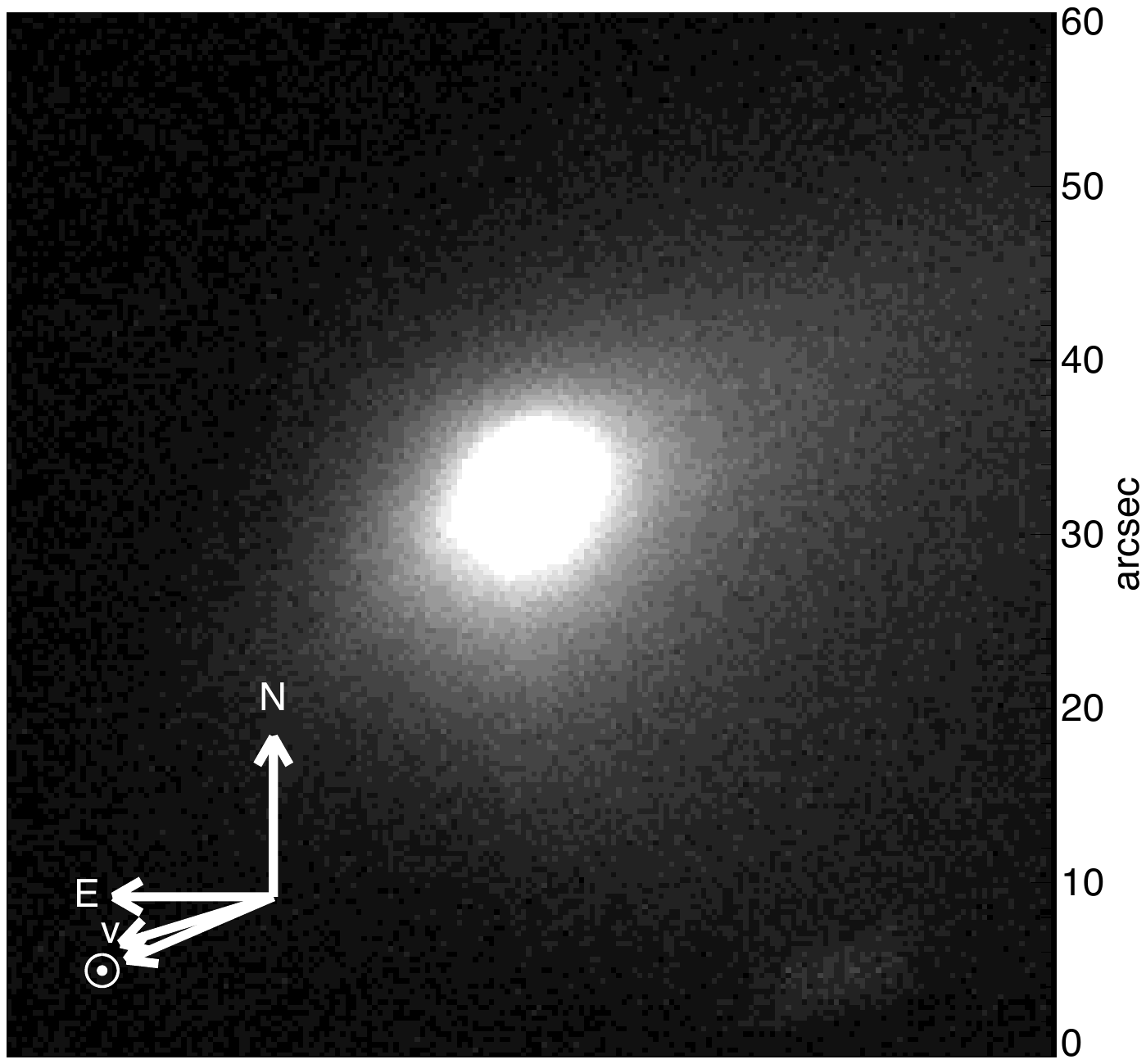}\put(5,85){\textcolor{white}{Oct 2015}}\end{overpic}\\
   \begin{overpic}[width=0.325\columnwidth,trim=111 230 111 230, clip]{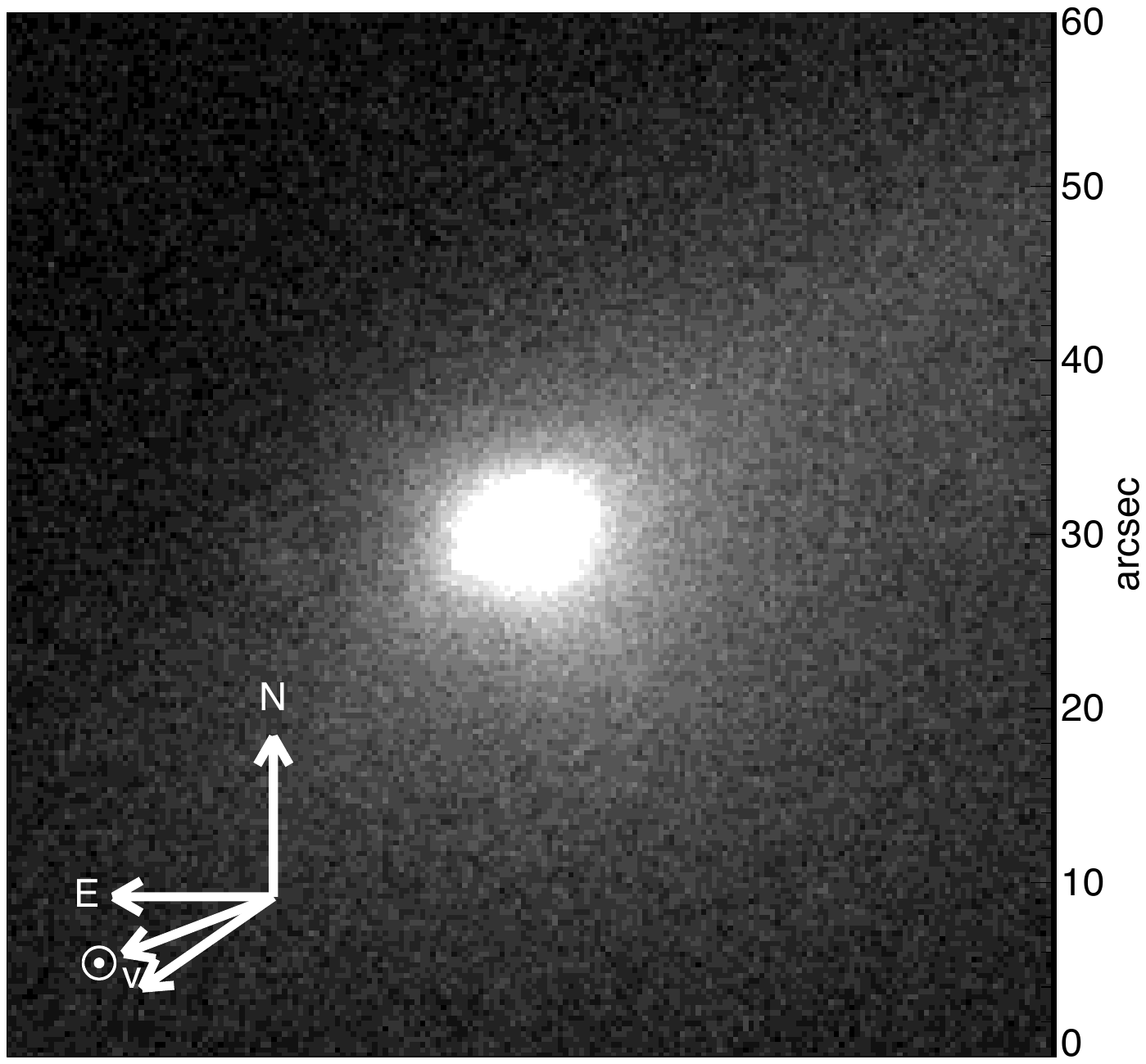}\put(5,85){\textcolor{white}{Jan 2016}}\end{overpic}
   \begin{overpic}[width=0.325\columnwidth,trim=111 230 111 230, clip]{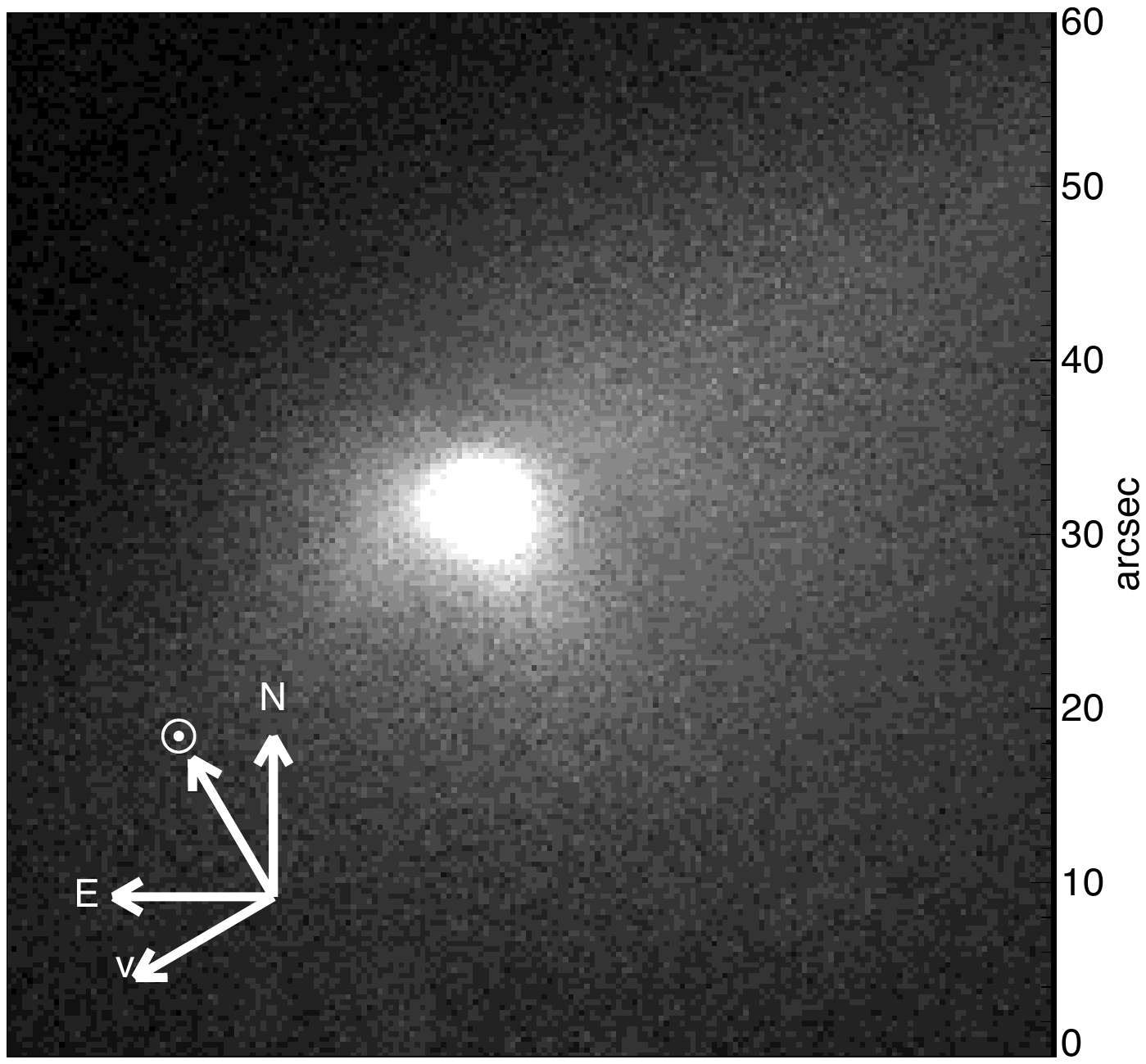}\put(5,85){\textcolor{white}{Mar 2016}}\end{overpic}
   \begin{overpic}[width=0.325\columnwidth,trim=111 230 111 230, clip]{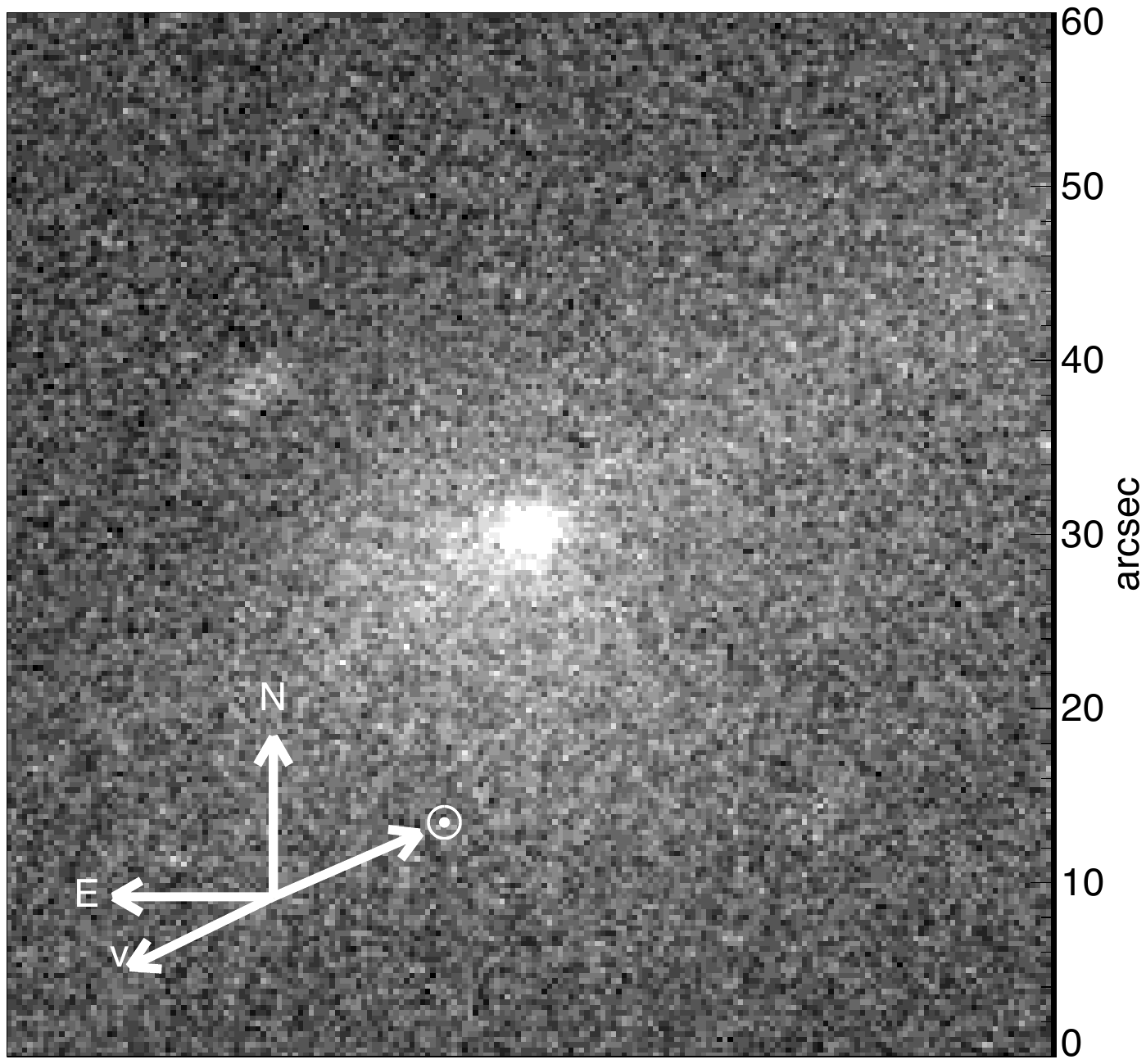}\put(5,85){\textcolor{white}{Jun 2016}}\end{overpic}\\
  
      \caption{$R$-band Images of the comet, 1 arcminute on each side. Arrows indicate the direction of the orbital velocity ($v$) and Sun ($\odot$) directions, i.e. opposite the expected direction of the dust trail and ion tail respectively. Image dates, telescopes and exposure times: 
      2014/02/27, VLT/FORS, 10x50s; 
      2014/07/01, VLT/FORS, 31x50s; 
      2014/10/22, VLT/FORS, 39x50s; 
      2015/05/21, VLT/FORS, 2x30s;
      2015/07/18, LT/IO:O, 10x20s;
      2015/10/07, LT/IO:O, 9x15s;
      2016/01/10, LT/IO:O, 3x120s;
      2016/03/10, LT/IO:O, 14x180s;
      2016/06/03, LT/IO:O, 3x180s.
      May 2015 image shows reflection from bright star out of FOV (above comet).
      }
         \label{fig:morphology}
   \end{figure}

\begin{figure}
\begin{center}
\includegraphics[width=\columnwidth]{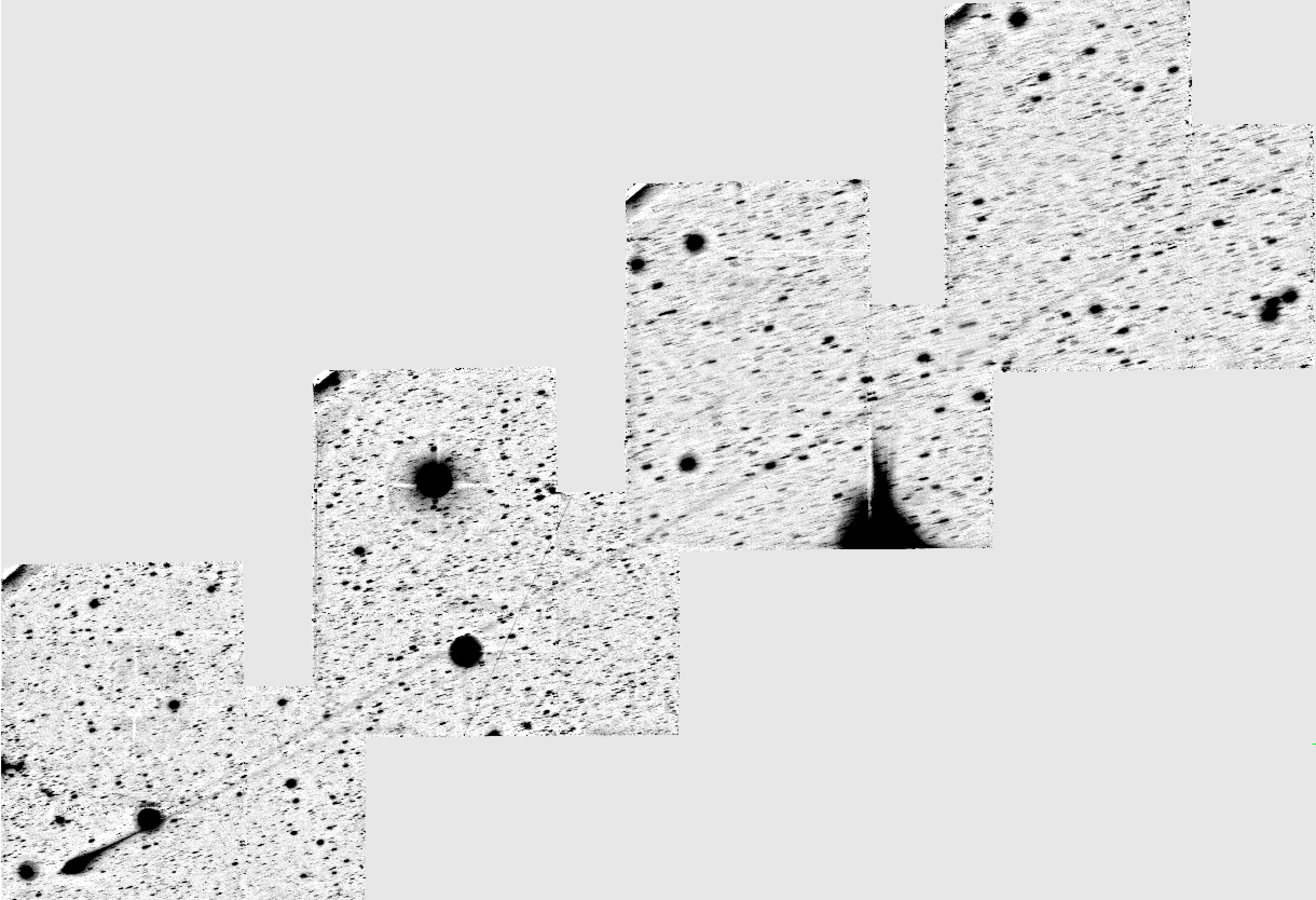}
\caption{Wide field image taken with the 2.5 m INT in March 2016, showing the long trail (approximately 2 degrees).}
\label{fig:INT}
\end{center}
\end{figure}

\begin{figure}
\includegraphics[width=0.32\columnwidth]{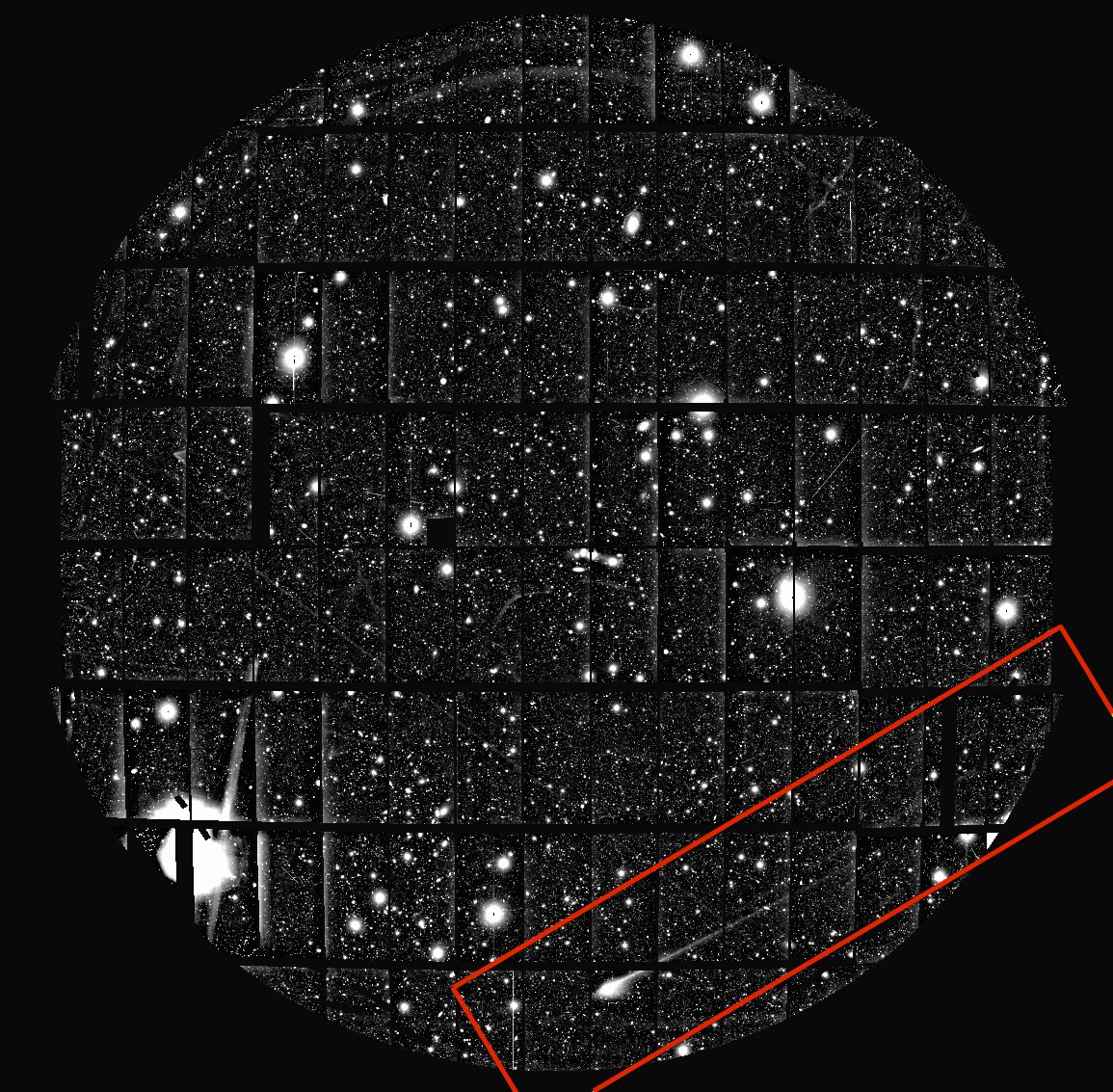}
\raisebox{0.9\height}{
\begin{tabular}{l}
\includegraphics[width=0.66\columnwidth]{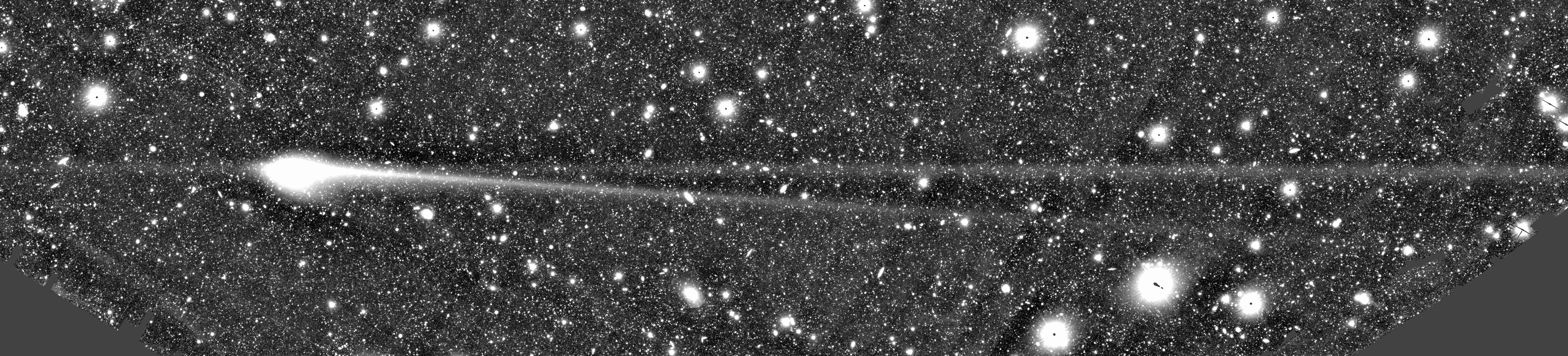}\\
\includegraphics[width=0.66\columnwidth]{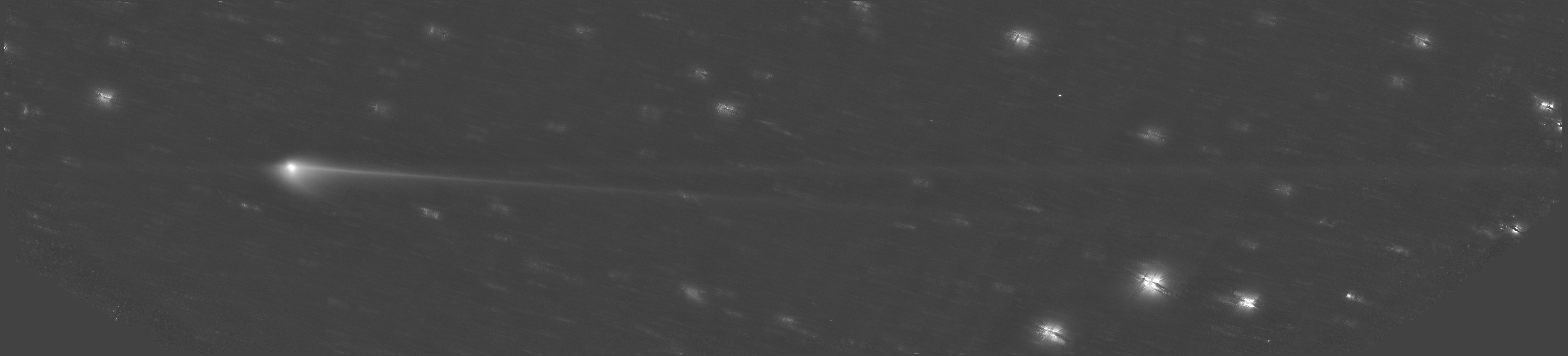}
\end{tabular}
}
\caption{Wide field image taken with the 8 m Subaru telescope and the Hyper Suprime-Cam, taken on 2016/03/08 when the comet was at 2.5 au from the Sun. Left: Full field of view of HSC (single frame), showing the region containing the comet. Top right: Extracted comet region (62.3 $\times$ 14.2 arcmin), total of 10 $\times$ 6 minute exposures stacked. Bottom right: same images median combined after shifting to account for comet motion.}
\label{fig:subaru}
\end{figure}

The appearance of the comet changed significantly during the campaign, as it went from inactive, through perihelion, and then as the activity faded as it retreated from the Sun. A selection of images that illustrate the general appearance of the comet is shown in fig.~\ref{fig:morphology}. 

The earliest observations in the campaign (in 2013 and early 2014) showed a point source, an apparently inactive nucleus, although photometry and observations from Rosetta/OSIRIS indicated that detectable activity began early in 2014, when the comet was more than 4~au from the Sun\cite{Snodgrass2016,Tubiana2015}. The comet became visibly active during 2014, showing a short tail at least 10 arcseconds long, corresponding to $25\,000$ km at the distance of the comet, by the time of the Philae landing in November.

When the comet was again visible from Earth in 2015 it was considerably brighter, with a tear-drop shape and  a long tail ($\sim$ 70" / $120\,000$ km), showing a similar appearance through perihelion. The apparent tail length increased as the comet continued to brighten, and also as it became visible in darker skies. As the comet began to retreat from the Sun it took on a distinct aspect, similar to that shown on previous orbits, with a broad coma and a clear narrow tail (possibly a so-called `neck-line', composed of dust released 180$^\circ$ in true anomaly before the date of observation), along with a very long dust trail tracing its orbit. It maintained this appearance until the end of the campaign, although fading as it reached $\sim$3.5~au from the Sun by the end of observations in 2016.

The long dust trail was particularly apparent in wide field images obtained in early 2016, when the comet was well placed for deep imaging. Figure~\ref{fig:INT} shows a mosaic taken with the wide field camera on the 2.5~m INT on La Palma, with each L-shaped field-of-view covering approximately half a degree on a side. The trail can be traced to over $10^7$~km from the comet, but is seen to be at a slightly different angle from the tail/neck-line feature that is brighter closer to the comet. 
The `two tails' (trail and neck-line) are also apparent in deep and wide-field images obtained serendipitously with the 8~m Subaru telescope on Mauna Kea, in observations on 2016/03/08 using the new Hyper Suprime-Cam, which is a mosaic imager with a 1.5~degree field of view (fig.~\ref{fig:subaru}). Detailed modelling of these structures is still to be done, but it is clear that the trail and neck-line become more apparent post-perihelion. Finson-Probstein models\cite{FinsonProbstein,Beisser1987} indicate that the narrow tail structure should contain old dust; for example in early November 2015 it should be dust that is at least 400 days old, released long before perihelion\cite{Boehnhardt-MNRAS}. 

It is worth noting that only dust tails (or trails) were observed. Despite
dedicated searches there was no ion tail feature seen. The observing geometry
(and the low inclination of 67P's orbit) meant that such observations were
always challenging, but it seems that any ion features near to the comet were
too faint to separate from the dust, and further away could not be detected
even in the deepest images. While the comet was relatively bright a number of
observations were made through narrowband filters, either special cometary
filters from the Hale-Bopp set\cite{Farnham2000}, or by selecting suitable
bandpasses from larger narrowband sets (e.g. for the 10~m Gran Telescopio Canarias [GTC]). At the 1~m Lulin Optical Telescope (LOT) in
Taiwan and at the 2.4~m at Lijiang in China, copies of the same narrowband
filter set flown on Rosetta/OSIRIS\cite{Keller-OSIRIS} were used to observe the
comet from the ground, providing a direct comparison between the inner 10s of
km seen from the spacecraft and the whole coma. Preliminary results from
narrowband imaging do not reveal obvious differences between gas and dust
morphologies, other than the large scale gas coma being more symmetrical with
no tail seen, but the observations generally have relatively poor S/N and analysis is ongoing.
Photometry from these images can be used to derive gas production rates, which
are consistent with spectroscopy results (see section~\ref{sec:spec}), and (in
the case of the LOT/Lijiang data) will be used to make direct comparisons with
the Rosetta/OSIRIS gas observations\cite{Lin-inprep,Bodewits2016}.


\begin{figure}
\begin{center}
\includegraphics[width=0.315\columnwidth]{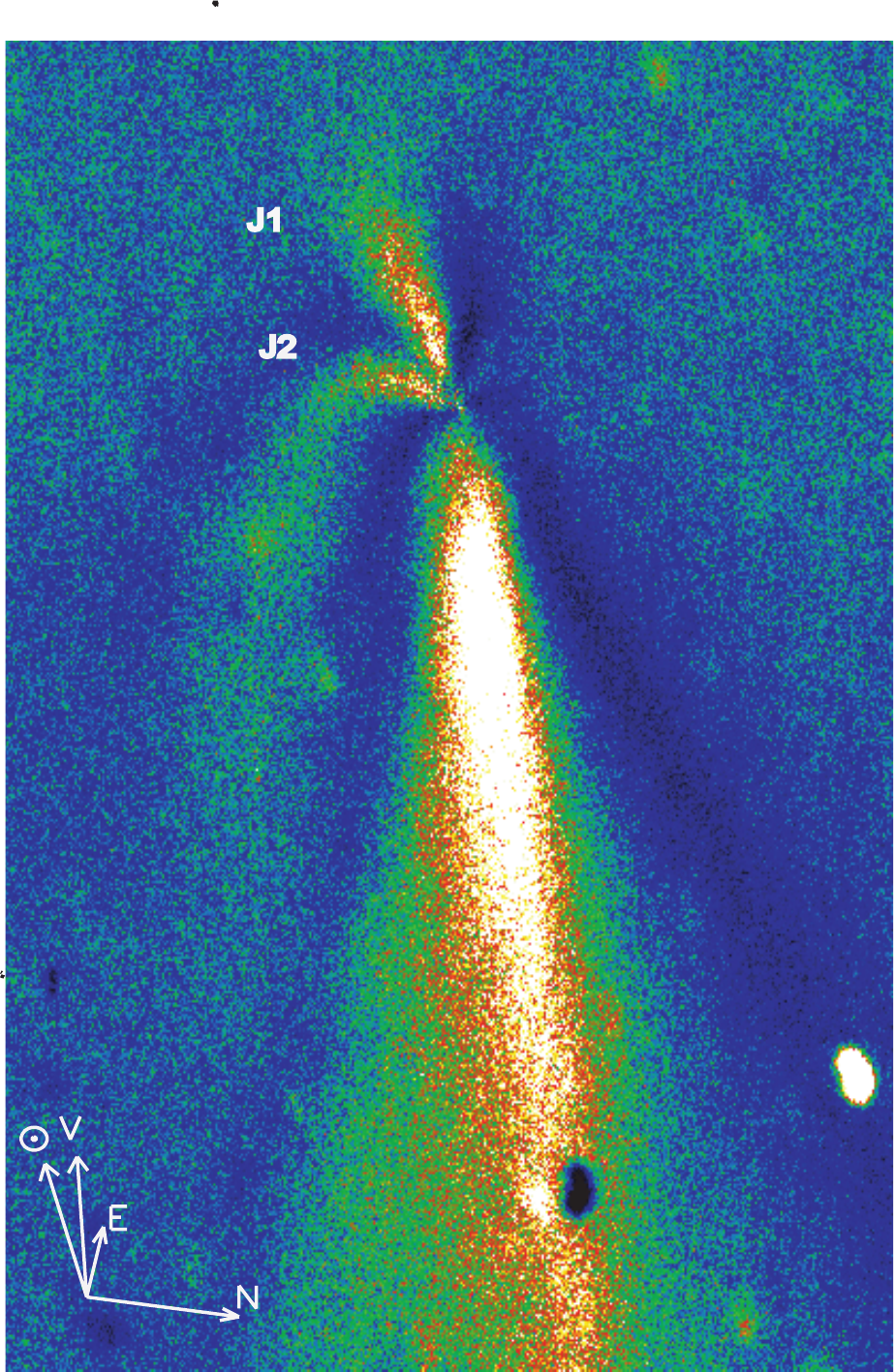}
\includegraphics[width=0.675\columnwidth]{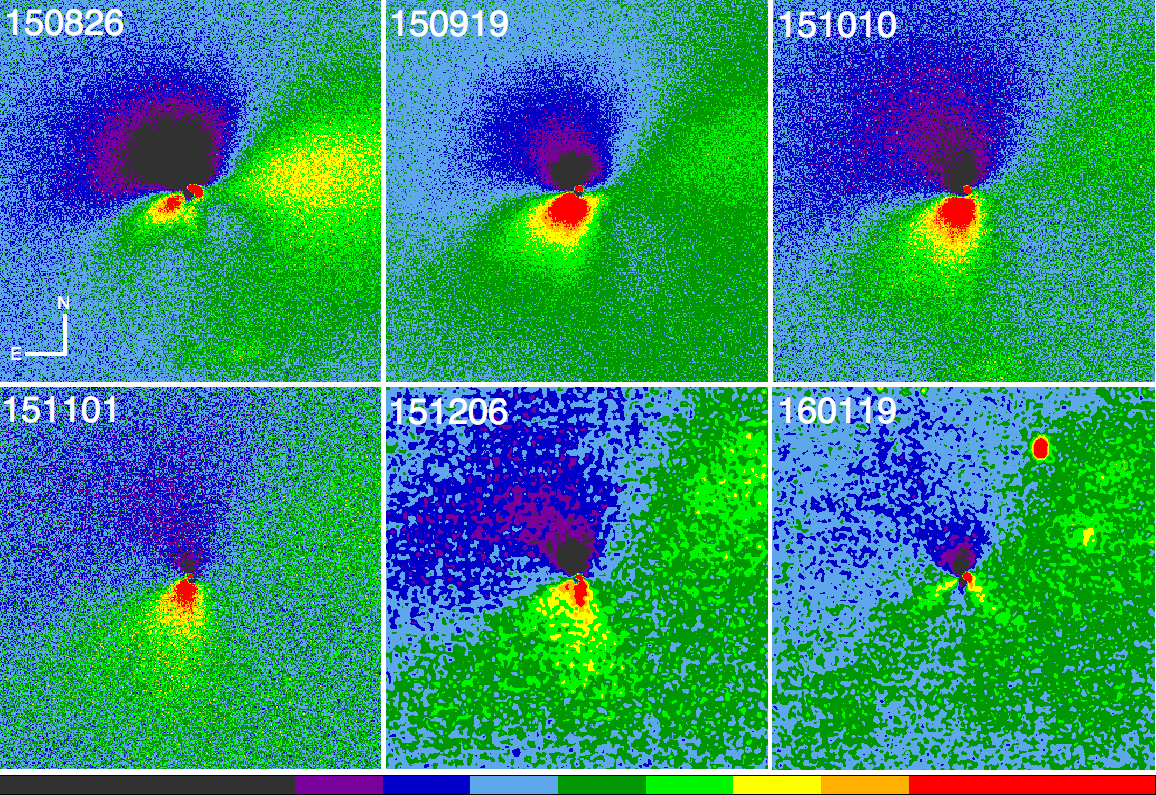}
\caption{{\bf Left:} `Jets' in the coma (labelled J1, J2), as seen from the 6 m BTA telescope of the SAO (Russia), on 2015/11/08. Image is $\sim$~100,000 km across. {\bf Right:} Enhanced Gemini NIRI $J$-band images of the comet monthly from August 2015
through January 2016 (date give as YYMMDD). Images are centred on the
comet and an azimuthal
median profile has been subtracted to reveal the fainter underlying
structure. At times (August,
January) two distinct structures can be discerned that match those labelled J1, J2 in the SAO image while at other times they overlap to appear as a single larger
structure towards the southeast. All images have the same colour scheme
with red/orange bright and blue/purple/black faint, but different
colour scales. Each image is 50,000 km on a side and has north up and
east to the left. The Sun and the direction of the comet's orbital velocity
are towards the southeast in all panels, and do not change significantly over this period (fig.~\ref{fig:morphology}). The red blob within a few pixels
of the centre in all panels is an artifact of the enhancement; trailed stars
can be seen as streaks in August, October, and January.}
\label{fig:jets}
\end{center}
\end{figure}

The morphology within the coma on $10^3$--$10^5$~km scales is more complex than the large scale tail/trails picture. Various image enhancement techniques can be used to reveal structure within the coma, and a similar pattern is visible in different data sets and using different techniques. A stable pattern of fans or jets is seen  using either Larson-Sekanina\cite{Larson+Sekanina84} processing or subtraction of an azimuthal median profile (fig.~\ref{fig:jets}). Although we refer to these structures as jets, they may projections of broader dust flows (fans), and do not necessarily relate to the narrow jets seen in Rosetta images of the inner coma\cite{Lin-AA}. 
This pattern showed a slow evolution\cite{Boehnhardt-MNRAS,Knight-inprep,Zaprudin-inprep}, with the relative intensity of the different jets approximately following the changing seasons on the comet -- the Southern structures are brighter around perihelion when this hemisphere of the nucleus is illuminated\cite{Keller2015}. Preliminary analysis of coma morphology seen throughout the apparition is consistent with predictions\cite{Vincent2013} for the source regions and pole solution [Vincent, priv. comm.], indicating that these jets are features that reappear each orbit\cite{Lara2015,Boehnhardt-MNRAS,Knight-inprep}.

Further analysis of the shape of the coma reveals evidence for a short-lived change (outburst) in late August 2015, around the time of peak activity post-perihelion, and a change in the slope of the coma profile indicating possible dust fragmentation\cite{Boehnhardt-MNRAS}. Observations with the HST revealed differences in  polarisation within the jets compared with the background coma\cite{Hadamcik-MNRAS}. Finally,  models  can be employed to recreate the coma morphology  based on assumptions on dust properties. In the case of 67P, where in situ instruments provide many constraints on these properties, detailed models have linked the 2014 observations and early OSIRIS observations\cite{Moreno2016a} and will further investigate the changing morphology with time\cite{Moreno2016b}.

\section{Composition}\label{sec:spec}

\begin{figure}
\begin{center}
\includegraphics[width=\columnwidth]{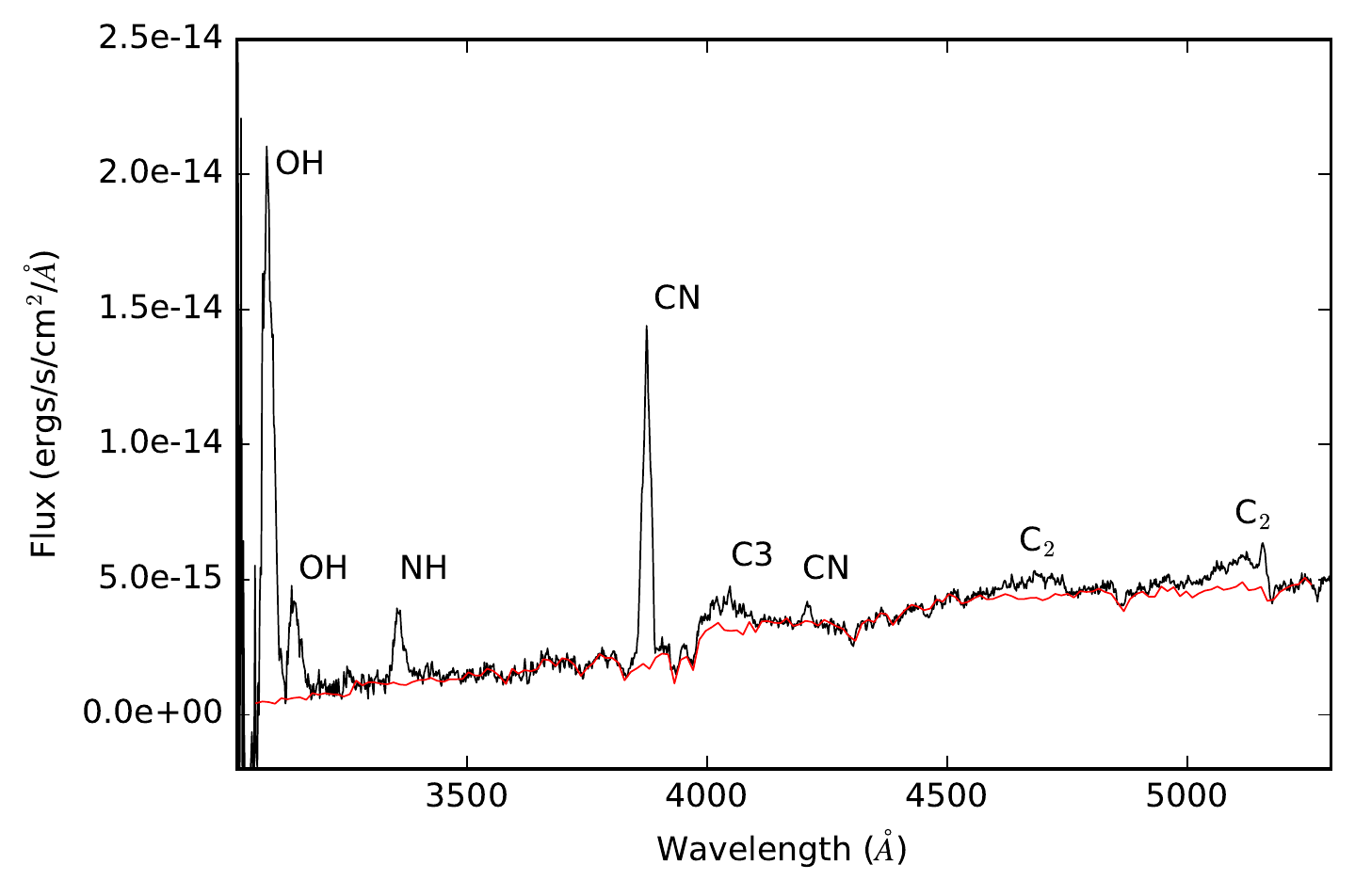}
\caption{Spectrum taken with the blue arm of ISIS on the WHT, on 2015/08/19, with the comet just past perihelion. The narrow red line shows a scaled solar analogue (i.e. the continuum/dust signal) for comparison. The strongest emission bands are identified.}
\label{fig:WHTspec}
\end{center}
\end{figure}

From the ground, the composition of comets (or any astrophysical object) is generally probed by spectroscopy. The solid components of the comet (its nucleus and dust coma) have similar and generally featureless spectra, reflecting sunlight back with a red slope in the visible range and a more neutral reflectance spectrum in the near-IR. The spectrum of the dust coma seen early in the mission matched Rosetta/VIRTIS observations of the nucleus\cite{Snodgrass2016,Capaccioni2015Sci}. As the comet approached perihelion Rosetta observations revealed some variation across the surface, including exposed ice patches. Near-IR spectroscopy with Gemini-N (GNIRS) and the NASA IRTF (SpeX) was obtained with the goal to look for and characterise the signatures of water-ice grains in the coma as previously obtained in the much more active comet 103P/Hartley 2 \cite{Protopapa2014}. Analyses of the GNIRS and SpeX data are ongoing \cite{Protopapa-inprep}.

The gas coma of comets is far more revealing, as emission features from various species can be measured across all possible wavelengths, for bright enough comets. 
Rosetta's own remote sensing spectrograph suite observed the gas coma from the UV through to the sub-mm (ALICE, VIRTIS, MIRO), detecting water already from June 2014 onwards and also mapping CO$_2$, OH and CN, among other species\cite{Gulkis2015Sci,BM-MNRAS,Migliorini2016}.
From the ground we were mostly limited to observations of 67P in the visible range, where emissions from so-called `daughter' species are seen (e.g. fig.~\ref{fig:WHTspec}). Using the detection of these species to probe the composition of the parent ices in the nucleus requires the use of photochemistry models, which Rosetta presents a unique opportunity to test (by comparing these observations with in situ measurements of parent gasses escaping directly from the nucleus). The spectrum in fig.~\ref{fig:WHTspec} was taken with the ISIS spectrograph on the WHT within a week of perihelion, and shows a fairly typical comet emission pattern, with obvious OH and CN bands, and weaker C$_2$, C$_3$ and NH features. The intensity of  C$_2$ and C$_3$  in spectra of 67P recorded in this campaign is relatively low (compared with the strong CN band), placing 67P in the carbon-chain depleted class of comet, in agreement with earlier observations\cite{Ahearn95}. 

\begin{figure}
\begin{center}
\includegraphics[width=\columnwidth]{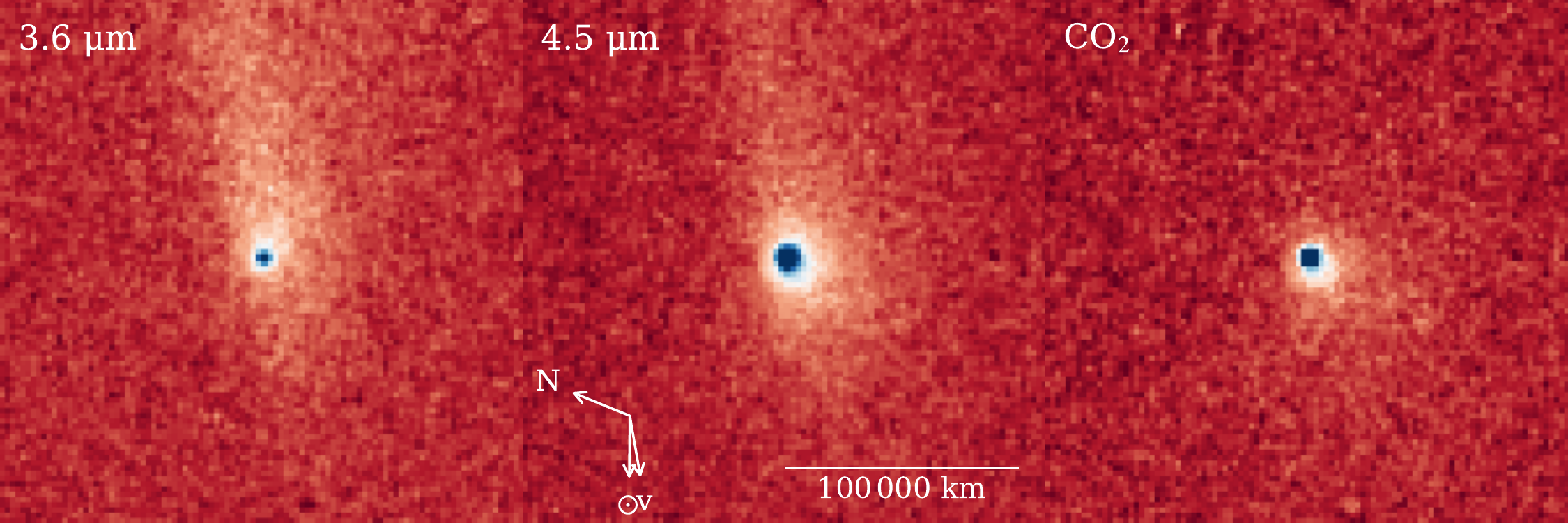}
\caption{Spitzer/IRAC images of the comet at 3.6 and 4.5~$\mu$m (left and center).  The CO$_2$ coma (right) is apparent after the 3.6~$\mu$m image (dust) is subtracted from the 4.5~$\mu$m image (dust and gas).  Celestial North (N), the projected orbital velocity ($v$), and the projected direction of the Sun ($\odot$) are marked with arrows.  Each image is approximately 200\,000~km on a side.}
\label{fig:spitzer-coma}
\end{center}
\end{figure}

Longer wavelength and/or higher resolution spectroscopy was possible when the comet was at its brightest. High-resolution near-IR spectroscopy, useful to separate cometary water emission lines from the terrestrial atmosphere, was attempted close to perihelion (2015/07/26-31) in good conditions with CSHELL on the IRTF, but resulted only in (3$\sigma$) upper limits of $Q({\rm H}_2{\rm O}) \le 5.1 \times 10^{27}$ molecules~s$^{-1}$ and $Q({\rm C}_2{\rm H}_6) \le 9.9 \times 10^{25}$ molecules~s$^{-1}$, assuming a rotational temperature of 40 K. These limits were close to the total water production interpolated from Rosetta results ($\sim 3.9 \times 10^{27}$ molecules~s$^{-1}$ for late July\cite{Fougere2016}), suggesting a detection was just out of reach.  
At longer IR wavelengths Spitzer/IRAC \cite{Werner2004, Fazio2004} 
 and WISE \cite{Mainzer2011}
photometry can be used to estimate the production rate of CO$_2$ \cite{Reach2013,Bauer2012,Bauer2015}, another major parent species in the coma.  The comet's CO$_2$-to-dust ratio at 2.8 to 3.0~AU (post-perihelion) appeared relatively low compared with other comets observed at similar distances in the same survey\cite{Kelley-inprep}.  Figure~\ref{fig:spitzer-coma} shows the four Spitzer epochs median combined into a single image.  An asymmetry in the 4.5~$\mu$m coma due to emission from the CO$_2$  $\nu_3$ band at 4.26~$\mu$m suggests the production of this gas is dominated by the southern hemisphere, similar to Rosetta/VIRTIS and Rosetta/ROSINA observations from elsewhere in the orbit \cite{Haessig2015Sci, BockeleeMorvan2015, Fink2016}.  Observations were also carried out in the sub-mm range, using the large ground-based facilities IRAM and ALMA to detect HCN (a  parent of CN) and CH$_3$OH at rates of $\sim 9 \times 10^{24}$ and $\sim 2 \times 10^{26}$ molecules~s$^{-1}$, respectively, in September 2015. From above Earth's atmosphere, observations with the Odin satellite in November 2015 searched for a water signature, but were only able to give upper limits ($Q({\rm H}_2{\rm O}) \le 3.3 \times 10^{27}$ molecules~s$^{-1}$), again close to the Rosetta value at that time\cite{Fougere2016}.

\section{Polarimetry}\label{sec:pol}

\begin{figure}
\begin{center}
\includegraphics[width=\columnwidth]{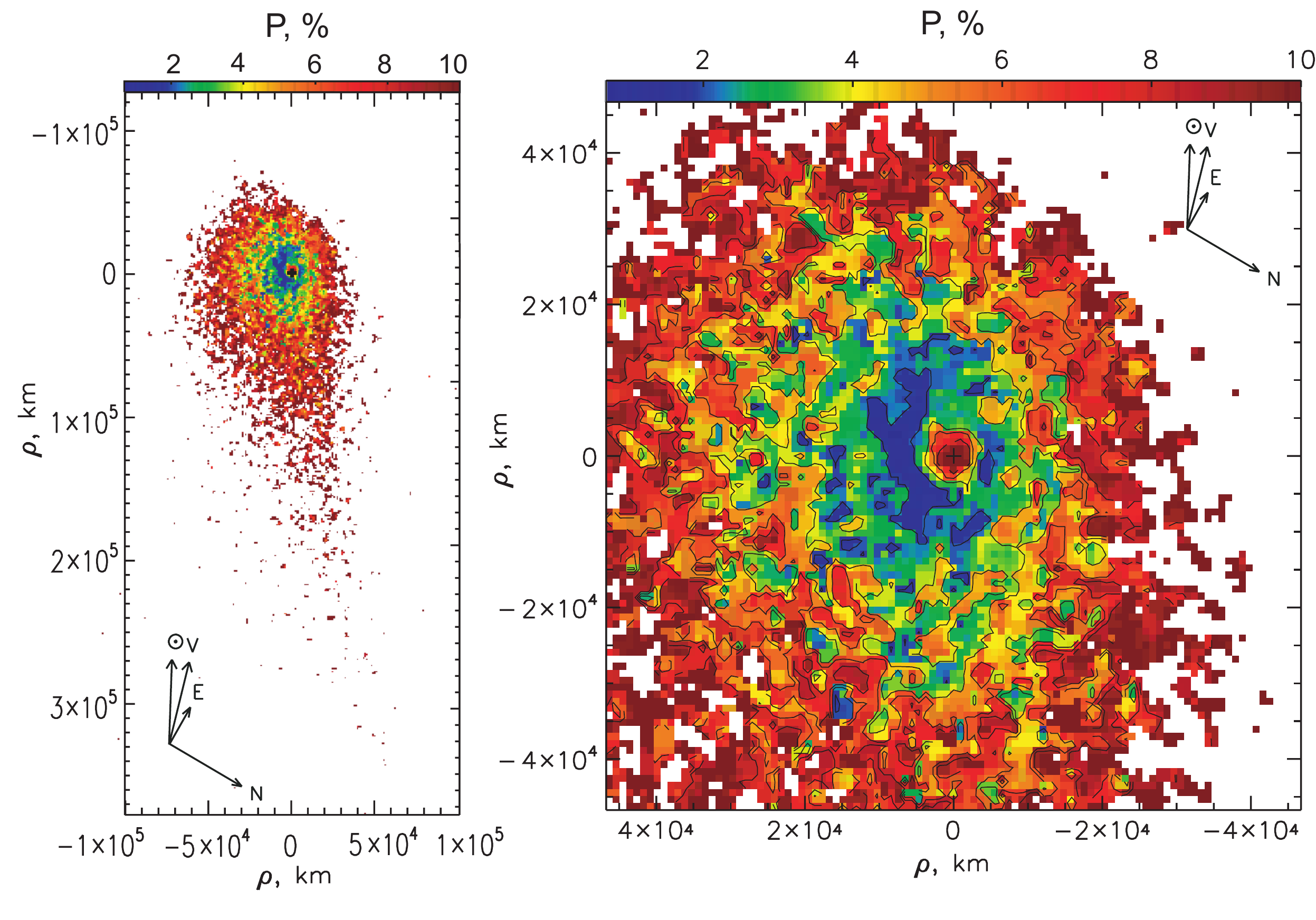}
\caption{Distribution of linear polarisation (P, \%) in comet 67P (left: coma and tail; right: zoom in on coma). Observation obtained with SAO 6~m telescope on 2015/11/08.}
\label{fig:pol}
\end{center}
\end{figure}

Polarimetry, and more specifically linear polarisation imaging and phase curves, provides evidence for changes in dust physical properties and gives clues to size and morphology of the dust particles (see, e.g., ref~\cite{Kolokolova2015}). 
Polarimetric images of 67P have been obtained from the HST ACS/WFC in 2014, 2015 and 2016. In August and November 2014, the comet, still far away from the Sun, was observed at low galactic latitudes; for a phase angle about 15.5$^\circ$, the average polarisation was nominal, about -2\% \cite{Hines+ACLR2016}. Three months after perihelion, in November 2015, the comet was still quite active, with conspicuous structures in intensity; for a phase angle about 33$^\circ$, the average polarisation was in agreement with what had been noticed at the previous passage \cite{Hadamcik10}, above average values, suggesting significant changes in the properties of dust aggregates ejected by the comet after perihelion \cite{Hadamcik-MNRAS,Hines-inprep}.
Polarimetric images have also been obtained from the VLT, the WHT, Rohzen observatory, and BTA, the 6~m Russian telescope, between August 2015 and April 2016. The polarisation maps (fig.~\ref{fig:pol}) provide evidence for different properties (e.g. size, shape, porosity) in the dust particles across the coma\cite{Ivanova-inprep}.


\section{Total activity}\label{sec:activity}

One of the fundamental measurements provided by the Earth-based view of 67P was an assessment of the `total' activity of the comet, which is an important reference for Rosetta results. Activity measurements from the spacecraft necessarily depend on various models, to reconstruct the global activity from a local measurement at one position inside the coma. Reassuringly, attempts to compare the measurements from various instruments with the ground-based total activity view produce largely consistent results (including between different Rosetta instruments, although there are some differences between ROSINA and VIRTIS), suggesting that the models used to interpret local measurements are valid\cite{Fougere2016,Hansen-MNRAS}.

The total activity of the comet can be measured in various ways from the ground, looking at the dust or gas coma. The total dust activity is easiest to follow, and can be assessed using broad-band photometry (typically $R$-band, to avoid contamination in the bandpass by gas emissions). Archival imaging was used to measure the total activity in previous orbits\cite{Snodgrass2013}, and make predictions for 2014-2016, and the same sort of measurements were applied to the campaign data: We measure the total brightness of the comet in $R$-band within a constant circular aperture of radius $\rho$ = $10\,000$ km at the distance of the comet. Some observations were taken with Johnson or Cousins $R$-band filters, while others used the $r'$ filter of the SDSS system (and VLT/FORS observations used the `R\_SPECIAL' filter that is somewhere between these). All photometry was calibrated onto the Cousins $R$ (Landolt) photometric scale, using transformations from the SDSS system and the $(g-r) = 0.62 \pm 0.04$ colour of the comet measured with the LT near perihelion where necessary\cite{Snodgrass-MNRAS}. This allowed direct comparison with predictions. 

\begin{figure}
\begin{center}
\includegraphics[width=0.7\columnwidth,trim=40 35 85 80,clip]{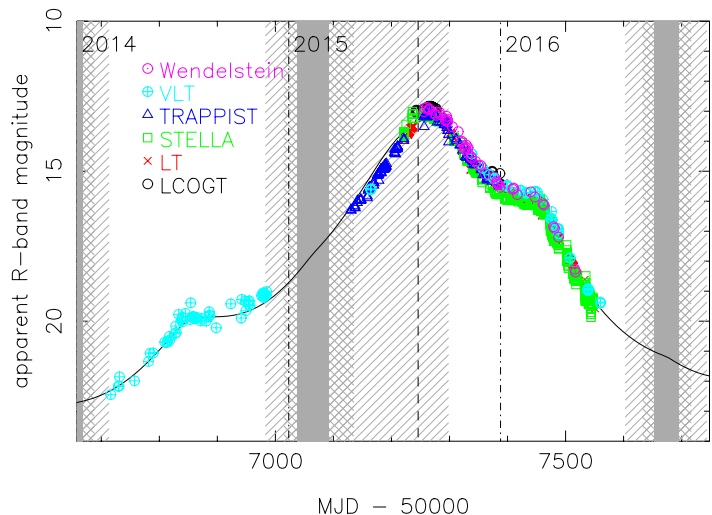}
\caption{Total $R$-band magnitude of the comet, compared with prediction from previous orbits (solid line). Solar elongation is indicated with hatching as in fig.~\ref{fig:visibility}.}
\label{fig:rmag}
\end{center}
\end{figure}

We show the measured total brightness of the comet in $R$-band in fig.~\ref{fig:rmag}, along with the prediction from three previous orbits. It is immediately clear that the comet's brightness followed the prediction very well, implying that the activity level of the comet is consistent from one orbit to the next. It is also clear that the campaign resulted in very complete coverage of the 2014-2016 period, with regular observations whenever it was possible to obtain them. The peak in activity in late August ($\sim$2 weeks after perihelion) is obvious. There are subtle differences between data sets taken with different filters ($R$ vs $r'$), implying a possible change of colour in the coma with time. The change in colour is small and has not yet been studied in detail, but likely relates to the changing gas production\cite{Snodgrass-MNRAS,Boehnhardt-MNRAS}.There are no large outbursts or other sudden brightness changes, and the phase function assumed in the prediction (a linear phase function with $\beta = 0.02$ mag deg$^{-1}$) clearly gives a decent fit over the range of phase angles seen from Earth ($\alpha \le 35^\circ$). The simple power law dependencies on heliocentric distance on which the predictions are based (flux $\propto r^{-5.2}$ pre-perihelion and $\propto r^{-5.4}$ post-perihelion\cite{Snodgrass2013}) can be used to give a good first order description of the dust brightness. Just before perihelion the observed brightness appears to be slightly below the prediction, but this is probably an effect of the peak in activity being offset slightly from perihelion, which isn't considered in the simple power law model\cite{Snodgrass-MNRAS}. In terms of the widely used $Af\rho$ parameter for quantifying cometary activity\cite{Ahearn84}, we find that the peak in activity was around $Af\rho \approx 1000$~cm. 
The Wendelstein data support an $Af\rho$ power law dependence on heliocentric distance with an exponent of -3.7 to -4.2, depending on the phase function assumed, using data from mid-September to the end of December 2015\cite{Boehnhardt-MNRAS}. This is close to the $Af\rho \propto r^{-3.4}$ post-perihelion from the prediction paper, and within the range of previous determinations discussed there\cite{Snodgrass2013}.

The total activity can also be measured in terms of gas production rates. The most abundant species released by the comet is water, but this is difficult to measure from the ground -- only comets considerably brighter  than 67P can be regularly observed with high resolution spectroscopy to separate cometary water emission lines (e.g. in the near-IR) from the terrestrial atmosphere. In the weeks pre-perihelion an upper limit of $Q({\rm H}_2{\rm O}) \le 5.1 \times 10^{27}$ molecules s$^{-1}$ was measured with IRTF, as described in section~\ref{sec:spec}. The alternative way to obtain water production rates from the ground is through observation of the daughter species such as OH, via the emission bands around 308 nm in the UV, or the [OI] lines near 630 nm. These are also challenging for a faint comet, given the strong absorption in the UV  by atmospheric ozone and the need for high resolution to separate oxygen lines from terrestrial ones. Successful detections of OH were made in 67P, primarily using the ISIS spectrograph on the WHT (fig.~\ref{fig:WHTspec}). A production rate of $Q({\rm OH}) = 2.6 \times 10^{27}$ molecules s$^{-1}$ was found on 2015/08/19 (within a week of perihelion), which corresponds to $Q({\rm H}_2{\rm O}) = 3.2 \times 10^{27}$ molecules s$^{-1}$. Further observations with ISIS were attempted until April 2016, but the OH production rate was only measurable relatively close to perihelion\cite{Fitzsimmons-inprep}. 
Observations of OH emission with the Lowell 1.1m and Kron photoelectric photometer (and narrowband filters) were used to derive water production rates of $Q({\rm H}_2{\rm O}) = 7.7$ and $3.4 \times 10^{27}$ molecules s$^{-1}$ on 2015/09/12  and 2015/10/15, respectively.
[OI] lines were detected using UVES on the VLT and HIRES on Keck in late 2015.  While used in the past as a reliable proxy for H$_2$O production in comets \cite{Morgenthaler2001,Fink2009,McKay2015}, the detection of abundant molecular oxygen in the coma of 67P \cite{Bieler2015} complicates interpretation of the observed [OI] line fluxes in terms of H$_2$O production rates.

\begin{figure}
\begin{center}
\includegraphics[width=0.7\columnwidth,trim=40 35 85 80,clip]{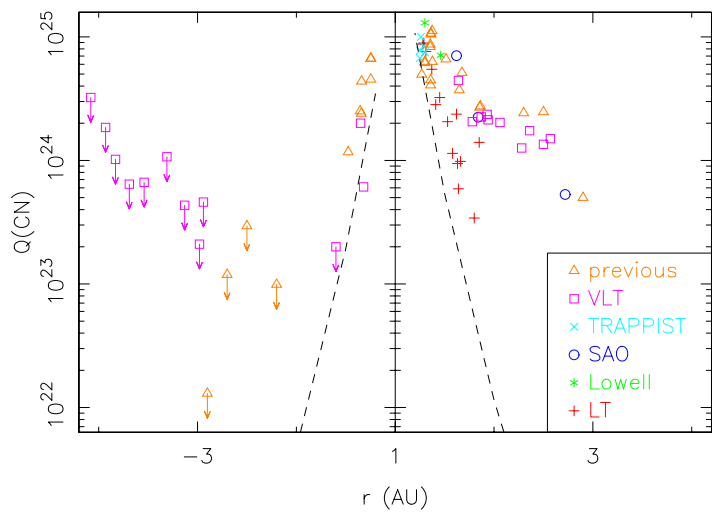}
\caption{CN production rate (in molecules s$^{-1}$) as a function of heliocentric distance (negative values indicate pre-perihelion data, positive post-perihelion).  We include data from previous orbits\cite{Schleicher2006,Schulz2004,Lara11,Guilbert2014}, narrowband photometry from TRAPPIST and the Lowell 1.1~m around perihelion, and spectroscopy from the VLT, SAO 6 m and LT (see key in plot for symbols). Error bars are not included for clarity. The dashed line shows a scaled version of the dust scaling law plotted in fig.~\ref{fig:rmag}. In general, pre-perihelion data are only upper limits (marked with arrows) until relatively low $r$, when the rate climbs quickly pre-perihelion, with a similar slope to the dust fit. CN emission can be detected to larger distance post-perihelion, with a shallow decrease in $Q$(CN). Data sets from different telescopes mostly agree, although the LT production rates post-perihelion are generally lower than those measured with larger telescopes, but with significant scatter. This behaviour appears to be seen in data from previous orbits too, with little change in total CN production (there appears to be a reasonable match between the `previous' points and those taken in this campaign).}
\label{fig:CNplot}
\end{center}
\end{figure}

The most easily detected gas species in a typical cometary coma is the CN radical, with a strong emission band around 389 nm, so the longer term gas production rate was monitored by observations of this band. Preliminary Rosetta results suggest that CN production does largely follow the water production rate, although there may be some long term variation in the relative proportions [Altwegg, priv. comm.]. The detection of CN was still challenging in 67P, with sensitive searches with the VLT and FORS in 2014 unsuccessful\cite{Snodgrass2016}. When the comet returned to visibility in 2015 CN was still not immediately detected, despite the considerably brighter coma, and an upper limit of $2\times10^{23}$ molecules~s$^{-1}$ was found in May with the VLT, even though the heliocentric distance was only 1.6 au. The rate then rapidly increased, with a positive detection finally achieved in early July at $2\times10^{24}$ molecules~s$^{-1}$ (1.35 au), and a range of facilities were able to make observations via spectroscopy or narrowband imaging in the months after perihelion, while VLT and SAO 6 m observations continued to trace CN out to $\sim$ 3 au post-perihelion. Estimates of the CN production rate against heliocentric distance are shown in fig.~\ref{fig:CNplot}. The strong asymmetry around perihelion is clear -- while detection was impossible with even the best telescopes until just before perihelion inbound, CN was measured for many months post-perihelion.

\section{Discussion and open questions}\label{sec:discussion}

The observing campaign largely demonstrated 67P to be a fairly typical Jupiter family comet, with a predictable and smoothly varying activity level, and no major outbursts or unusual events. In this way it confirmed that Rosetta was seeing typical behaviour of a typical object -- an important statement that allows the conclusions from Rosetta measurements to be taken as generally true for comets. However, there were some surprises that require further investigation. The most obvious of these is the puzzling difference between the symmetrical rise and fall of total activity as measured by dust brightness, and the sharp onset and then slow decrease in activity measured by CN gas production. Taken at face value this implies a significant change in dust-to-gas ratio with time, but the observation does not agree with the symmetrical rise and fall in total gas production seen by in situ Rosetta instruments\cite{Fougere2016,Hansen-MNRAS}. While Rosetta/ROSINA sees some change in the CN/water ratio with time, these are subtle, and in general in situ measurements find that the CN and water production rates appeared to be correlated (while the relative abundance of other major species to water, e.g. CO$_2$/water, varies across the nucleus and with time\cite{Haessig2015Sci,BM-MNRAS}). The apparent turn on and off of CN as seen from Earth are near to the dates of equinox on the comet, so this could be a seasonal effect (i.e. the CN parent is mostly released from the Southern hemisphere), but this was not obvious in Rosetta/ROSINA measurements [Altwegg, priv. comm.]. This implies a  difference between in situ and whole coma measurements, which still needs to be explained. One possibility is a distributed  source of CN, at distances $> 100$ km from the nucleus, that is not seen by Rosetta. A more detailed analysis of the long term gas and dust production rate monitoring will appear in a future paper\cite{Opitom-inprep}. 

If one of the conclusions of the parallel Earth and Rosetta observations is that the bright CN band cannot be used as a reliable tracer of total gas production, an equally important test will be to see how well more direct tracers compare with in situ measurements. Although more difficult to perform, observations of OH are generally thought to have the advantage of having a single well known parent (water) and therefore tracing total (water dominated) gas production directly. In the few months post-perihelion where we could perform OH measurements from the ground, Rosetta production rates are based on models that extrapolate to the whole coma from a local (or single line of sight) measurement. These models agree with ground-based photometry for (scaled) dust production\cite{Hansen-MNRAS}, but have not yet been directly compared with ground-based gas measurements. A further complicating factor is the discovery, from Rosetta/ALICE, that the dissociation models used to get daughter species fluxes need to take into account electron impact as well as Solar UV radiation\cite{Feldman2015}. This has been taken into account in studies of the gas production via Rosetta/OSIRIS narrowband imaging of the inner coma\cite{Bodewits2016}, but the implications for the larger scale coma need to be considered. 

Finally, while the total brightness evolution of 67P was very smooth, there is evidence of short term variations (i.e. outbursts). Outbursts from comets vary in scale from the frequent but small scale events seen as Deep Impact approached comet 9P/Tempel 1\cite{Ahearn-DI, Feldman2007}, to events that can cause the coma to brighten by many magnitudes (such as the mega-outburst of 17P/Holmes in 2007). The abundant photometry on 67P from this campaign will allow careful searches for small outbursts, and in particular tests to see if the many short-lived events seen as bright jets in Rosetta imaging\cite{outbursts} are correlated with changes in the total brightness. One potentially significant outburst, in late August 2015, has already been identified from ground-based data due to the effect it had on the overall shape of the inner coma\cite{Boehnhardt-MNRAS}. There is also a possible signature in ground-based photometry of the outburst seen by many of Rosetta's instruments in February 2016\cite{Gruen-outburst}, although the comet was close to the full moon at that time, and also near to opposition and therefore phase function effects on the total brightness need to be carefully considered.

\section{Conclusion}

An unprecedented long-term campaign of observations followed comet 67P throughout the Rosetta mission, including characterisation of the nucleus and activity levels before the spacecraft arrived, and parallel to its operational period (2014-2016). This made 67P one of the best studied short period comets, with observations following it from its inactive state through perihelion and back out to beyond 3~au, despite challenging geometry (low solar elongation) for much of this apparition. The parallel observations with the long-term in situ monitoring from Rosetta provide a unique opportunity to test observational techniques and models against `ground-truth'. We find that the comet's brightness largely varies in a smooth and predictable way, with no major outbursts or changes from orbit-to-orbit, but subtle variations can be identified. The morphology of both the inner coma and the large scale tails and trails is also repeatable between orbits, and implies a stable pattern of activity, which we hope to correlate with the detailed view of active regions seen by the spacecraft. The comet's composition is typical of the carbon-depleted class, but the dust and CN gas production rates varied in different ways around perihelion, indicating possible differences in composition across the nucleus. With $\sim$~1300 hours of observation over 4 years, there is a wealth of ground-based data to compare with the treasure trove of Rosetta results: A large number of detailed follow up studies are on going, and will be published in the coming year(s). 
\vskip6pt

\enlargethispage{20pt}


\dataccess{It is our intention that all observational data from the campaign will be archived alongside Rosetta instrument data at the ESA Planetary Science Archive (\url{http://www.cosmos.esa.int/web/psa}). In addition, much of the raw data are (or will be) available from individual observatory archive facilities. Observing log information available at \url{http://www.rosetta-campaign.net} will also be permanently archived at the PSA.}

\aucontribute{C. Snodgrass coordinated the campaign and drafted the manuscript. All authors contributed to observations and/or data reduction, and read and approved the manuscript.}

\competing{The authors declare that they have no competing interests.}

\funding{C. Snodgrass is supported by an UK Science \& Technology Facilities Council (STFC) Rutherford fellowship.}

\ack{We acknowledge the contribution of the Europlanet EU FP7/H2020 framework in supporting meetings that initiated this campaign in 2012 and brought many of us together to discuss the results in 2016. We thank the Royal Society and the organisers of the `Cometary science after Rosetta' meeting for the invitation to present the results of the campaign. 
S.F. Green acknowledges support from the STFC (Grant ST/L000776/1).
J. Kleyna is supported by NSF grant 1413736.
H.J.~Lehto, B. Zaprudin and A. Somero  acknowledge  the  support  of  the Academy of Finland (grant number 277375).
J. Licandro and J. de Le\'on acknowledge support from the AYA2015-67772-R  (MINECO, Spain).
J.~Lasue and A.C. Levasseur-Regourd acknowledge support from CNES, the French Space Agency, for this work in relation with CONSERT and MIDAS on board Rosetta.
Z.Y. Lin and X. Wang were supported by NSC 102-2112-M-008-003-MY3 of the Ministry of Science and Technology, Taiwan, and National Natural Sciences Foundations of China (contract No. 11073051 and No. 11473066).
C. Opitom acknowledges the support of the FNRS.
Finally, we thank the many observatories involved in these observations for their support in allocating significant time to observing 67P, especially ESO, Gemini, and Observatorios de Canarias del IAC (through the CCI International Time Programme) for enabling the long-term baseline of observations.
We are grateful for the efforts of various support astronomers in assisting with the observations: In particular
Ian Skillen at the ING; 
Fumiaki Nakata, Finet Francois, and the HSC Queue Working Group at Subaru;
Thomas Granzer at STELLA;
David Abreu and Pablo Ruiz from Ataman Science, Spain, for supporting the OGS observations.
}



\end{document}